# A Kushner-Stratonovich Monte Carlo Filter Applied to Nonlinear Dynamical System Identification

S Sarkar[1], S R Chowdhury[1], M Venugopal[2], R M Vasu[2] and D Roy[1*]

[1]Computational Mechanics Lab, Department of Civil Engineering, Indian Institute of Science, Bangalore

[2]Department of Instrumentation and Applied Physics, Indian Institute of Science, Bangalore

[*]Corresponding Author; Email: royd@civil.iisc.ernet.in

**Abstract:** A Monte Carlo filter, based on the idea of averaging over characteristics and fashioned after a particle-based time-discretized approximation to the Kushner-Stratonovich (KS) nonlinear filtering equation, is proposed. A key aspect of the new filter is the gain-like additive update, designed to approximate the innovation integral in the KS equation and implemented through an annealing-type iterative procedure, which is aimed at rendering the innovation (observation-prediction mismatch) for a given time-step to a zero-mean Brownian increment corresponding to the measurement noise. This may be contrasted with the weight-based multiplicative updates in most particle filters that are known to precipitate the numerical problem of weight collapse within a finite-ensemble setting. A study to estimate the *a-priori* error bounds in the proposed scheme is undertaken. The numerical evidence, presently gathered from the assessed performance of the proposed and a few other competing filters on a class of nonlinear dynamic system identification and target tracking problems, is suggestive of the remarkably improved convergence and accuracy of the new filter.

## 1. Introduction

Stochastic filters, as a modern tool for dynamic system identification of interest across a broad range of areas in science and engineering, involve estimating the dynamically evolving states (processes) and/or model parameters conditioned on an experimentally observed noisy data set of known functions of the process variables till the current time. Within a complete probability space $(\Omega, \mathcal{F}, P)$, equipped with an increasing filtration $\{\mathcal{F}_t, 0 \leq t \leq T\}$ consisting of $\sigma$-subalgebras of $\mathcal{F}$, following are the generally adopted forms of the process and observation models, typically represented as Ito stochastic differential equations (SDEs):

$$dX_t = b(X_t,t)dt + f(X_t,t)dB_t \tag{1.1}$$

$$dY_t = h(X_t,t)dt + dW_t \tag{1.2}$$

Here $X_t := X(t) \in \mathbb{R}^n$ is the hidden process state, which is only partly revealed by the noisy observation process $Y_t := Y(t) \in \mathbb{R}^q$ generating the sub-filtration $\mathcal{F}_t^Y$. $b: \mathbb{R}^n \times \mathbb{R}^+ \mapsto \mathbb{R}^n$ obtains the non-linear drift term in Eq. (1.1). The diffusion matrix $f: \mathbb{R}^n \times \mathbb{R}^+ \mapsto \mathbb{R}^{n \times m}$ and the $m$-dimensional standard $P$-Brownian motion $B_t \in \mathbb{R}^m$ together determine the process noise. $h: \mathbb{R}^n \times \mathbb{R}^+ \mapsto \mathbb{R}^q$ is the non-linear observation function and $W_t \in \mathbb{R}^q$ a $q$-dimensional zero-mean $P$-Brownian motion representing the observation noise. It is assumed that the conditions [1] for the existence of weak solutions to the above SDEs are satisfied. A stochastic filter then aims at obtaining the conditional (filtering) distribution of, say, a scalar-valued function $\phi(X_t)$, $\phi \in C_b^2$ (the set of bounded and twice continuously differentiable functions), given the observation history $\{Y_s \mid s \in (0,t]\}$. Thus the estimate $\pi_t(\phi)$ is defined through the measure-valued process $\pi_t$ that is measurable with respect to the observation process; i.e.:

$$<\pi_t, \phi> = \pi_t(\phi) := E_P\left(\phi(X_t) \mid \mathcal{F}_t^Y\right) \tag{1.3}$$

Except for a few special cases wherein the Kalman-Bucy filter [2] yields 'exact' closed-form solutions to the filtering problem involving linear drift and observation functions along with additive Gaussian process and observation noises, the estimates are usually obtained by approximate analytical schemes like the extended Kalman filter (EKF) [3], the unscented Kalman filter (UKF) [4] or, more appropriately, via sequential Monte Carlo (SMC) techniques [5-7], which often use an ensemble of weighted realizations (called particles) of $X_t$ thus providing an empirical approximation to the filtering distribution. There are numerous schemes that approximate the conditional distribution of system states, the evolution of which is described by the nonlinear filtering equations. A survey of such numerical schemes in the context of nonlinear filtering may be found in [8]. Unfortunately, most such SMC techniques are plagued with the problem of 'particle impoverishment', especially for higher dimensional problems wherein the weights progressively tend to a point mass. Here the filtering scheme fails to provide any non-trivial updates to $X_t$, as obtained through Eq. (1.1), upon conditioning on $\mathcal{F}_t^Y$. Numerical evidence suggests that the typical ensemble size preventing 'weight collapse' increases exponentially with increasing system dimension [9]. Among the numerous research articles aiming at improving these SMC techniques, implicit sampling [10], improved resampling [11] and Markov chain Monte Carlo (MCMC) sampling based particle filters [12] are a few that have drawn attention.

The problem of nonlinear filtering could be solved if it were possible to approximate the solution of the Zakai equation that describes the evolution of the unnormalized conditional density of the system state. In this direction, [13, 14, 15, 16] have tried to approximate the conditional distribution using time and space discretizations or through functional series. For instance, the approximation of the conditional density using multiple Wiener/ Stratonovich integrals (MWI/MSI) have been derived and the error bounds involved in truncating the MWI/MSI series to a finite number of terms obtained in [13, 17]. In [18, 19], the authors have also numerically validated their proposed methods in approximating the Zakai equation via low-dimensional problems. Unfortunately, although the Zakai's equation is linear and has been widely studied, it has some serious deficiencies in numerical computations [20] which its nonlinear counterpart, the Kushner-Stratonovich (KS) equation circumvents.

The KS equation, the parent filtering equation derivable through Ito's expansion of the Kallianpur-Striebel formula, gives the evolution of $\pi_t(\phi)$ via a stochastic integral expression. However, owing to the moment closure problem for nonlinear, non-Gaussian dynamical systems, the KS equation cannot generally be reduced to stochastic PDEs for $\pi_t(\phi)$ so that they could be numerically integrated. In fact, attempts at numerically approximating the solution of the KS equation (e.g. a direct Euler-type discretization) do not generally yield an accurate and robust scheme. Indeed, particle based simulations in most SMC methods, e.g. the weighted particle system, may be thought of as Monte Carlo approximations to the KS equation using a conditional Feynman-Kac formula [21]. Most of these methods are however not free from the scourge of weight collapse, especially for larger filter dimensions.

We propose a novel particle based approach that closely mimics evolutions of the estimates through the KS equation and implements a nonlinear gain-like particle update, which is additive in nature and hence eliminates particle weighting-branching operations [22]. Moreover, by way of maximally utilizing the information available with the current observation (in a sense made precise later), the proposed time-recursive scheme crucially utilizes an inner iteration over every time-step, wherein an artificially introduced scalar diffusion multiplier associated with the innovation process is lowered over successive iterations as the estimate progresses towards the actual (i.e. the one corresponding to the solution of the KS equation).

The rest of the paper is organized as follows. Section 2 elaborates the proposed filtering methodology. We also provide a step-by-step algorithm of the proposed filter and a theorem on the order of convergence of the filter (due to approximations in time and over a finite ensemble) in Section 2. Section 3 presents a few numerical illustrations and this is followed by the concluding remarks in Section 4. The proof of the theorem (Theorem 1 in Section 2) is provided in Appendix I.

## 2. Methodology

Given an ordering $0 = t_0 < t_1 ... < t_i < ... < t_N = T$ of the time axis of interest, the estimate $\pi_t(\phi)$ of $\phi(X_t)$ over a generic time step $t \in (t_i, t_{i+1}]$ satisfies the KS equation:

$$\pi_t(\phi) = \pi_{t_i}(\phi) + \int_{t_i}^t \pi_s(\mathrm{L}_s(\phi)) ds + \sum_{\varsigma=1}^q \int_{t_i}^t \left\{ \pi_s\left(\mathrm{M}_s^\varsigma(\phi)\right) - \pi_s\left(h^\varsigma(\cdot, s)\right) \pi_s(\phi) \right\} dI_s^\varsigma \qquad (2.1a)$$

$I_t = \{I_t^\varsigma\} := \{Y_t^\varsigma - \pi_t(h^\varsigma(X, t))\} \in \mathbb{R}^q$ denotes the innovation process vector and $I_t^\varsigma$ its $\varsigma^{th}$ element. However, an important ingredient of the development to follow is a simplification of the second term on the RHS of the KS equation (2.1a) given by:

$$\int_{t_i}^t \pi_s(\mathrm{L}_s(\phi)) ds \approx \int_{t_i}^t \pi_i(\mathrm{L}_s(\phi)) ds = \int_{t_i}^t E_P(\mathrm{L}_s(\phi) | \mathcal{F}_i^Y) ds, \qquad (2.1b)$$

where $\pi_i(.) := \pi_{t_i}(.)$ and $\mathcal{F}_i^Y := \mathcal{F}_{t_i}^Y$. This approximation helps uncoupling the prediction and updating stages in the proposed filter over $t \in (t_i, t_{i+1}]$, an aspect that is found to be of numerical expedience during the initial phase of time evolution. By way of a 'maximal' assimilation of $Y_t$, the current observation, the present aim is to drive $\Delta I_i := I_{i+1} - I_i$ to a Brownian increment at the end of the filtering step over $(t_i, t_{i+1}]$, where $I_i := I_{t_i}$. Moreover,

$$L_t(\phi(x)) := \frac{1}{2}\sum_{\xi=1}^{n}\sum_{\eta=1}^{n} a^{\xi\eta}(x,t)\frac{\partial^2 \phi(x)}{\partial x^\xi \partial x^\eta} + \sum_{\xi=1}^{n} b^\xi(x,t)\frac{\partial \phi(x)}{\partial x^\xi}, \quad x = \{x^1,...,x^n\} \in \mathbb{R}^n \tag{2.1c}$$

and $M_t^\varsigma(\phi(x)) := h^\varsigma(x,t)\phi(x)$. Here $a := ff^T$ with $a^{\xi\eta}$ denoting the $(\xi,\eta)^{\text{th}}$ element of the matrix $a$. Similarly, $b^\xi$ is $\xi^{\text{th}}$ element of the vector $b$ and $h^\varsigma$, $Y_t^\varsigma$ are the $\varsigma^{\text{th}}$ elements of vectors $h$ and $Y_t$ respectively. Eq. (2.1a) is arrived at after averaging over the diffusion paths corresponding to the process noise $B_t$. Moreover, the first two terms on the RHS of Eq. (2.1a) recover Dynkin's formula for the predicted mean $E_P(\phi(X_t) | X(t_i) := X_i)$ according to the process dynamics of Eq. (1.1). By way of motivating the proposed KS filter, a particle based representation of Eq. (2.1a) may be conceived of by putting back, in the prediction component, the diffusion term for the process dynamics (an Ito integral with respect to $B_t$). In the updating stage, as the current observation $Y_t$ is available, the innovation vector $I_t$ may be treated as a pseudo-Markov process $I_t(\vartheta)$ in an artificially introduced time-like parameter $\vartheta$ and the aim is to drive $\Delta I_t(\vartheta) := I_t(\vartheta) - I_i$ weakly to $\Delta W_t$, the measurement noise increment, via inner recursions over $\vartheta$ for $t \in (t_i, t_{i+1}]$, often at $t = t_{i+1}$. In order to boost the mixing property of the associated transition kernel, the $\vartheta$-recursion, also referred to as the inner iteration, is accomplished by multiplying the innovation integral (the last term on the RHS of Eqn. (2.1a)) by $(1+\beta(\vartheta))$, where $\beta(\vartheta)$ is a scalar 'annealing-type parameter' (ATP) that is made to approach zero with progressing iterations so as to ensure consistency with the original form of the KS equation. It may be noted that in a class of filtering strategies, e.g. the so-called iterated filtering [23], iterations have been used to maximize the likelihood function for estimating the latent variables. Owing to the lack of dynamics, the latent variables are propagated via a perturbation parameter which is similar to our ATP in a sense that both are introduced for a better exploration of the search space. While in iterated filtering the problem is posed as one in stochastic optimization requiring several passes of a filter over the entire time window of interest, the scheme proposed herein aims at temporally localized iterations so designed as to satisfy the nonlinear filtering equation (or its time-discrete equivalent) over each time step.

*Prediction*

Consistent with the simplified form of the KS equation as above, the prediction SDE for $\phi(X_t)$, enabling particle-based simulation, is obtainable through Ito's formula applied to $\phi(X_t)$ where $X_t$ follows SDE (1.1). The integral form of the prediction equation over $t \in (t_i, t_{i+1}]$ is:

$$\phi(X_t) = \phi(X_i) + \int_{t_i}^{t} L_s(\phi(X_s)) ds + \int_{t_i}^{t} \nabla \phi(X_s) \cdot \{f(X_s, s) dB_s\} \tag{2.2}$$

$\{\nabla \phi(x)\}^\xi := \dfrac{\partial \phi(x)}{\partial x^\xi}$, $x = \{x^1, ..., x^n\} \in \mathbb{R}^n$ is the $\xi^{\text{th}}$ element of the gradient vector $\nabla \phi(x)$ and • denotes the scalar (dot) product of two vectors. The integrals in Eqn. (2.2) may be approximately evaluated, in a strong or weak sense, by any available numerical scheme [24, 25] thus producing the predicted ensemble $\{\phi(X_{i+1}^{(j)})\}_{j=1}^{N}$, $N$ being the ensemble size. Specifically, by choosing an appropriate set of such scalar functions $\{\phi^\xi(x) = x^\xi : \xi \in [1, n]\}$, one gets the ensemble of predicted states $\{X_{i+1}^{(j)}\}_{j=1}^{N}$.

*Iterated updates*

The iterative update equation for the KS filter based on a $\vartheta$-parameterization, presently realized through the discrete sequence $\{\vartheta_k : k = 0, ..., \kappa - 1\}$ (with $\vartheta_{k+1} > \vartheta_k$ for all $k$), at $t = t_{i+1}$ is given by:

$$\phi(\hat{X}_{i+1,k+1}) = \phi(X_{i+1}) + (1+\beta_k)U_{i+1,k} \qquad (2.3a)$$

where $\hat{X}_{i+1,k+1} := \hat{X}_{i+1}(\vartheta_{k+1})$ and $\beta_k = \Delta\vartheta_k = \vartheta_{k+1} - \vartheta_k$. An alternative form of this update, wherein the initial update corresponding to $k=0$ and $\beta_0 = 0$ is added to the prediction term before the subsequent updates with $\beta_k \neq 0$ for $k > 0$ take effect, is given by:

$$\phi(\hat{X}_{i+1,k+1}) = \phi(\hat{X}_{i+1,1}) + (1+\beta_k)U_{i+1,k}; \quad k = 1,\ldots,\kappa-1 \qquad (2.3b)$$

Here $\phi(\hat{X}_{i+1,1}) = \phi(X_{i+1}) + U_{i+1,0}$

and

$$U_{i+1}(\vartheta_k) := U_{i+1,k} = \sum_{\varsigma=1}^{q} \left\{ \pi'_{i+1}\left(M^{\varsigma}_{i+1}\left(\phi(\hat{X}_{i+1,k})\right)\right) - \pi'_{i+1}\left(h^{\varsigma}\left(\hat{X}_{i+1,k},t_{i+1}\right)\right)\pi'_{i+1}\left(\phi(\hat{X}_{i+1,k})\right)\right\}\Delta R^{\varsigma}_{i+1,k}$$

where $k = 0,\ldots,\kappa-1$, $\Delta R^{\varsigma}_{i+1,k} := \left\{\Delta Y^{\varsigma}_i - h^{\varsigma}\left(\hat{X}_{i+1,k},t_{i+1}\right)\Delta t_i\right\}$, $\Delta Y^{\varsigma}_i = Y^{\varsigma}(t_{i+1}) - Y^{\varsigma}(t_i)$ and $\hat{X}_{i+1,0} := X_{i+1}$. $\hat{X}_{i+1,k+1}$ denotes the $(k+1)^{th}$ inner-iterated update of $X_{i+1}$ conditioned on $\mathcal{F}^{Y}_{i+1}$ with the conventions that $X_{i+1} := \hat{X}_{i+1,0}$ and $\hat{X}_{i+1} := \hat{X}_{i+1,\kappa}$. $\pi'_{i+1}$ denotes the ensemble approximation to $\pi_{i+1}$ at time $t_{i+1}$ and $\Delta t_i = t_{i+1} - t_i$, i.e $\pi'(.) = (1/N)\sum_{j=1}^{N}(.)^{(j)}$. In Eq. (2.3b), $\beta_k := \vartheta_{k+1} - \vartheta_k > 0$ for $k > 0$ is the $\vartheta$-varying ATP which is analogous to temperature in a simulated annealing (SA) scheme [26]. Unlike the SA where the temperature is recursively reduced to zero whilst evolving a single Markov chain, the sequence $\{\beta_k \mid \beta_{k+1} \leq \beta_k\}$ is used in the KS filter to evolve an ensemble of $N$ pseudo-chains in $\vartheta$ so that, for a given $t$, the chains proceed in a controlled way to finally arrive at an ensemble that drives the pseudo-process $U_t(\vartheta)$ to a zero-mean Brownian increment corresponding to the measurement noise. The incorporation of $\beta_k$ in the update equation should therefore be supplemented with an appropriate choice of $\beta_1 \gg 0$ to begin the inner iterations and, for a given filtering problem, $\beta_1$ should typically be arrived at through a few trial runs of the filter. However, a more insightful, if not numerically expedient, choice could be based on the fact that a higher $\beta_1$ must necessarily correspond to a higher observation error $Y_{i+1} - h(\hat{X}_{i+1,1}, t_{i+1})$. One could thus artificially magnify this error by a scalar multiplier $\alpha \gg 1$ and finally obtain $\beta_1$ by adopting a minimization of the following scalar valued functional:

$$\beta_1 = \alpha \left\{ \arg\min_{\rho} \left( \left\| \Delta Y_i - h(\pi'(\hat{X}_{i+1,1}) + (1+\rho)U_{i+1,1}, t_{i+1})\Delta t_i \right\|_2 \right) \right\}$$

where $\|.\|_2$ is the Euclidean vector norm. The sequence $\{\beta_k\}$, whose elements for $k$ close to one are relatively higher, helps drive the predictions in the non-Newton (derivative-free) direction given by the vector $U_{i+1,k}$. The choice of the annealing-like schedule, ensuring $\beta_k \to 0$ as $k \to \infty$, is governed by two important factors, namely the computational speed and an effective exploration of the state space to attain the desired solution. A conservative schedule, useful for SA schemes involving a single Markov chain, is given by [27] $\frac{1}{\beta_{k+1}} = \frac{1}{\beta_k} + \frac{\lambda \rho(\beta_k)}{2\sigma^3(\beta_k)}$, where $\lambda < 1$ is a user-defined parameter. $\rho(\beta_k) := E\left[ \left( I_{i+1,k}^T I_{i+1,k} - I_{i+1,k-1}^T I_{i+1,k-1} \right)^2 \right]$ is the variance of an incremental energy-like term and $\sigma(\beta_k)$ is the standard deviation of $I_{i+1,k}^T I_{i+1,k}$, where $I_{i+1,k} := I_{i+1}(\vartheta_k)$. Although this schedule requires a large number of inner iterations to reduce $\beta_k$ to zero, it does improve results for the parameter estimation problems considered in Section 3. Nevertheless, since an ensemble-based formulation provides an additional means of exploring the phase space of $X_t$ through the particles, a more non-conservative schedule, e.g. $\beta_{k+1} = \frac{\beta_k}{\exp(k+1)}, k = 1,...,\kappa-1$ (which may not even qualify as a strict annealing schedule), appears to be more appropriate for the KS filter. Ideally, an appropriate stopping criterion should be used to fix $\kappa$ so as to ensure that $U_{i+1,k}$ is indeed a zero-mean discrete Brownian motion (random walk) for $k \geq \kappa - 1$. Towards this, one may employ the Kolmogorov-Smirnov test [28] in order to assess if the ensemble of realized observation errors $\left\{ Y_i - h(\hat{X}_{i+1,\kappa}^{(j)}, t_{i+1}) : j \in [1, N] \right\}$ indeed correspond to the known density of the random variable $W_{i+1}$, the observation noise at $t_{i+1}$. However, since the number of iterations required to pass this test could be computationally prohibitive, the numerical illustrations in this work are based on a fixed value of $\kappa$ assigned in a problem specific manner through a few trial runs whilst satisfying the constraint $\beta_{\kappa-1} \approx 0^+$. The typical value of $\kappa$ is around 10 in the numerical examples that we have reported in this work.

A further modification in the above particle based scheme, once the inner iterations are over, may be effected by way of relaxing the approximation (2.1b) to the second term on the RHS of the KS equation (2.1a). In doing this, final filtered state $\hat{X}_{i+1} := \hat{X}_{i+1,\kappa}$ is given by:

$$\phi\left(\hat{X}_{i+1}\right) = \phi\left(\hat{X}_i\right) + \frac{1}{2}[L_i(\phi(\hat{X}_i)) + L_{i+1}\left(\phi\left(\hat{X}_{i+1,\kappa-1}\right)\right)]\Delta t_i + \nabla\phi\left(\hat{X}_i\right)\bullet\{f\left(\hat{X}_i,t_i\right)\Delta B_i\} + U_{i+1,\kappa-1} \quad (2.4a)$$

On the other hand, if one were to iteratively correct the second term (on the right hand side of the above equation) using the inner updates $\hat{X}_{i+1,k}$, $k < \kappa$, large random fluctuations would possibly occur in $L_{i+1}(\phi)$, the term that encapsulates the physical laws governing the system dynamics, owing to the typically large values of $\beta_k$ for small $k$. This is especially true for small $i$, i.e. during the initial stages of time evolution. This would render the numerical solutions prone to overflows and thus destroy the filtering accuracy. Another way of avoiding such fluctuations would be to start updating the so-called prediction term only for $k \geq k_{max}$ (with $k_{max} < \kappa$) such that $\beta_k << 1$ for $k \geq k_{max}$. In that case, one may write the updates corresponding to the inner iterations for $k \geq k_{max}$ as:

$$\phi\left(\hat{X}_{i+1,k+1}\right) = \phi\left(\hat{X}_{i+1,k}\right) + (1+\beta_k)U_{i+1,k} \quad (2.4b)$$

For $k < k_{max}$, Eq. (2.3b) will remain applicable. Indeed the form of Eq. (2.4b) is used whilst proving the convergence of the inner iteration step in Theorem 1.

That the solution via the KS filter converges approximately to that of the KS equation may be intuitively demonstrated as follows. Using Eq. (2.4a) with $\beta_\kappa \approx 0$ and upon ensemble averaging over process noise, one may write as $\Delta t_i \to 0$:

$$\pi'_{i+1}(\phi(\hat{X}_{i+1})) \approx \pi'_i(\phi(\hat{X}_i)) + \pi'_{i+1}(L_{i+1}(\phi(\hat{X}_{i+1,\kappa-1})))\Delta t_i + \pi'_{i+1}(\nabla\phi(\hat{X}_i)\cdot\{f(\hat{X}_i,t_i)\Delta B_i\}) + \pi'_{i+1}(U_{i+1,\kappa-1})$$
(2.5)

where $\Delta B_i = B_{i+1} - B_i$ and

$$\pi'_{i+1}(U_{i+1,\kappa-1}) = \sum_{\varsigma=1}^{q}\left\{\pi'_{i+1}\left(M^{\varsigma}_{i+1}(\phi(\hat{X}_{i+1,\kappa-1}))\right) - \pi'_{i+1}\left(h^{\varsigma}(\hat{X}_{i+1,\kappa-1},t_{i+1})\right)\pi'_{i+1}\left(\phi(\hat{X}_{i+1,\kappa-1})\right)\right\}\Delta S^{\varsigma}_{\kappa-1}$$

with $\Delta S^{\varsigma}_{\kappa-1} = \left\{\Delta Y^{\varsigma}_i - \pi'_{i+1}\left(h^{\varsigma}(\hat{X}_{i+1,\kappa-1},t_{i+1})\right)\Delta t_i\right\}$. The term $\pi'_{i+1}(\nabla\phi(\hat{X}_i)\cdot\{f(\hat{X}_i,t_i)\Delta B_i\})$ is the expectation of an explicit Euler-approximation to the Ito integral $\int_{t_i}^{t_{i+1}}\nabla\phi(X_s)\cdot f(X_s,s)dB_s$, which is a zero-mean martingale and hence vanishes reducing Eq. (2.5) to:

$$\pi'_{i+1}(\phi(\hat{X}_{i+1})) \approx \pi'_i(\phi(\hat{X}_i)) + \pi'_{i+1}(L_{i+1}(\phi(\hat{X}_{i+1,\kappa-1})))\Delta t_i + \pi'_{i+1}(U_{i+1,\kappa-1})$$
(2.6)

Eq. (2.6) is indeed a discrete and empirical approximation to Eq. (2.1a), the original KS equation.

The algorithm (pseudo-code) for implementing the proposed filter is given below. For clarity of exposition, we consider $\phi(x) = x$.

1. Discretize the time interval of interest, say $[0,T]$, using a partition $\{t_0,t_1,...,t_M\}$ such that $0 = t_0 < t_1 < ... < t_M = T$ and $t_{i+1} - t_i = \Delta t$ uniformly for $i = 0,...,M-1$. Choose an ensemble size $N$.

2. Generate the ensemble of initial conditions $\{X_0^{(j)}\}_{j=1}^{N}$ for the state vector. For each discrete time instant $t_{i+1}, i = 0,...,M-1$ the following steps are carried out.

3. (Prediction:) Using $\{\hat{X}_i^{(j)}\}_{j=1}^{N}$, the update available at the last time instant $t_i$ (with the convention that $\{\hat{X}_0^{(j)}\}_{j=1}^{N} = \{X_0^{(j)}\}_{j=1}^{N}$), propagate each particle to the current time instant $t_{i+1}$ using an explicit Euler-Maruyama (EM) approximation to Eqn. (1.1), i.e.

$$X_{i+1}^{(j)} = \hat{X}_i^{(j)} + b(\hat{X}_i^{(j)}, t_i)\Delta t + f(\hat{X}_i^{(j)}, t_i)\Delta B_i, \ j = 1,..., N$$

4. (Initial update:)

$$\hat{X}_{i+1,1}^{(j)} = X_{i+1}^{(j)} + G_{i+1}\left(\Delta Y_i - h\left(X_{i+1}^{(j)}\right)\Delta t\right)$$

Evaluate each column of the $G_{i+1}$ matrix, i.e. $g^l := G_{i+1}(:,l) \quad l = 1, 2,..., q$ as,

$$g^l = \left(\frac{1}{N}\sum_{j=1}^{N} h\left(X_{i+1}^{(j)}(l)\right) X_{i+1}^{(j)}\right) - \left(\frac{1}{N}\sum_{j=1}^{N} h\left(X_{i+1}^{(j)}(l)\right)\right)\left(\frac{1}{N}\sum_{j=1}^{N} X_{i+1}^{(j)}\right)$$

$$G_{i+1} = \left[g^1, g^2,..., g^q\right]$$

Here $X_{i+1}^{(j)}(l)$ denotes $l^{\text{th}}$ element of the vector $X_{i+1}^{(j)}$.

5. (Iterated updates:) Set $k = 1$ and select $\beta_1$, $\kappa$, $k_{\max}$. Then update each particle as

$$\hat{X}_{i+1,k+1}^{(j)} = \hat{X}_{i+1,k}^{(j)} + (1+\beta_k)G_{i+1,k}\left(\Delta Y_i - h\left(\hat{X}_{i+1,k}^{(j)}\right)\Delta t\right)$$

However, during the initial stages of inner iterations (i.e. for $k \leq k_{\max}$), particles may be updated via the following map to avoid possible numerical oscillations in the updated solution:

$$\hat{X}_{i+1,k+1}^{(j)} = \hat{X}_{i+1,1}^{(j)} + (1+\beta_k)G_{i+1,k}\left(\Delta Y_i - h\left(\hat{X}_{i+1,k}^{(j)}\right)\Delta t\right)$$

Evaluate each column of the $G_{i+1,k}$ matrix, i.e. $g^l := G_{t_{i+1},k}(:,l), \ l = 1, 2,..., q$ as:

$$g^l = \left(\frac{1}{N}\sum_{j=1}^{N} h\left(\hat{X}_{i+1,k}^{(j)}(l)\right) \hat{X}_{i+1,k}^{(j)}\right) - \left(\frac{1}{N}\sum_{j=1}^{N} h\left(\hat{X}_{i+1,k}^{(j)}(l)\right)\right)\left(\frac{1}{N}\sum_{j=1}^{N} \hat{X}_{i+1,k}^{(j)}\right)$$

$$G_{i+1,k} = \left[g^1, g^2,..., g^q\right]$$

6. Set $k = k+1$. If $k < \kappa$, set $\beta_{k+1} = \frac{\beta_k}{\exp(k+1)}$, and go to step 5.

Else if $i < M$, go to step 3 with $i = i+1$, else terminate the algorithm.

Note that the inner iteration to implement the nonlinear additive update, shown through steps 4 and 5 in the pseudo-code, is inspired by and interpretable as a variant of stochastic Picard's iteration – a tool often used by mathematicians to prove existence and uniqueness of solutions to nonlinear SDEs. In other words, the proposed inner iterations are a means to asymptotically secure the true solution of the KS equation modulo the time-discretization and sampling errors. Indeed, the uniqueness of the true solution (i.e. the conditional distribution) corresponding to the filtering problem described by the KS equation (2.1a) is ensured under the somewhat strong assumption of uniform Lipschitz continuity of the drift and diffusion coefficients via the work of Kurtz and Ocone [34], who pose the problem as one of filtered martingale and prove uniqueness by extending the original theory of Stroock and Varadhan. Now, it may be worthwhile to take stock of the approximations involved in arriving at the filtered estimate. First, the explicit EM method used to generate the predicted solution provides a source for integration errors whose weak local order is $\mathcal{O}(\Delta t)$ [35]. Additionally, integration errors also accrue owing to the iterative updates, which are consistent with the EM-based treatment of a diffusion term. Finally, one must also account for the MC error arising due to the empirical representations of the associated probability distributions over a finite ensemble. It is therefore of interest to obtain some formal estimates of the errors involved on these counts.

The *a-priori* error is presently estimated by considering the two approximations separately as

$$\left(E_P\left[\left|\pi_{i+1}(\phi)-\pi_{i+1}^{\prime e}(\phi)\right|^2\right]\right)^{\frac{1}{2}} \leq \left(E_P\left[\left|\pi_{i+1}(\phi)-\pi_{i+1}^{e}(\phi)\right|^2\right]\right)^{\frac{1}{2}} + \left(E_P\left[\left|\pi_{i+1}^{e}(\phi)-\pi_{i+1}^{\prime e}(\phi)\right|^2\right]\right)^{\frac{1}{2}} \quad (2.7)$$

where the superscript $e$ denotes the EM-based approximation. The first term on the right hand side represents the mean-square error in approximating the conditional distribution at $t_{i+1}$ in time alone and the second term obtains the MC error (due to an ensemble approximation) following time discretization. The error estimate for the first term in Eqn. (2.7) is obtained on the same lines as in [21]. The second term is bounded by following steps similar to those given in [29]. We assume $n=q=1$ only for expositional convenience, even though the error orders indicated below remain unchanged otherwise.

***Theorem 1***:

Let $\phi \in C_b^2(\mathbb{R})$. Assume that there exist constants $M_1, M_2 > 0$ such that

$$|b(t,x)-b(t,y)|+|f(t,x)-f(t,y)| \leq M_1|x-y| \qquad (2.8)$$

$$|b(t,x)|+|f(t,x)| \leq M_2|1+|x|| \qquad (2.9)$$

and

$$E_P\left[|X_0|^2\right] < \infty \qquad (2.10)$$

Assume additionally that $h$ is a bounded and continuous function. Furthermore, we assume that $\phi$ is sufficiently smooth so that $\phi(x)$ and its derivatives satisfy an inequality of the form

$$|\phi(x)| \leq M_3\left(1+|x|^a\right) \qquad (2.11)$$

for positive constants $M_3, a$. Then there exist constants $D' > 0$ and $D'' > 0$, independent of $\Delta t$, such that

$$\left(E_P\left[|\pi_{i+1}(\phi) - \pi_{i+1}^{\prime e}(\phi)|^2\right]\right)^{\frac{1}{2}} \leq D'(\Delta t)^{\frac{1}{2}}$$

$$+ \frac{D''}{\sqrt{N}} \begin{pmatrix} (3|\Delta Y_{i+1}|+5\|h\|)\|h\|\|\phi\| + (|\Delta Y_{i+1}|\Delta t)\|L(h\phi)\| \\ +(|\Delta Y_{i+1}|\Delta t\|h\|+(\Delta t)^2\|h\|^2)\|L(\phi)\| \\ +(|\Delta Y_{i+1}|\Delta t\|\phi\|+4(\Delta t)^2\|h\|\|\phi\|)\|L(h)\|+\Delta t\|L(\phi)\|+\|\phi\| \end{pmatrix}$$

$$+ \left(\|\mathbf{P}^1(x,|\phi|^2)\|+\|\mathbf{P}^\kappa(x,|\phi|^2)\|\right)d_{Hell}(\mathbf{P}^1(x,\cdot),\mathbf{P}^\kappa(x,\cdot))$$

$$(2.12)$$

where $\mathbf{P}^k(x,\cdot)$ is defined as the transition law corresponding to the additive update equation at the $k^{\text{th}}$ inner iterate. $\|\cdot\|$ denotes the supremum norm and $d_{\text{Hell}}(\cdot,\cdot)$ is the Hellinger metric.

The proof is provided in Appendix I.

## 3. Numerical Illustrations

For purposes of numerical demonstration of the filter performance, we consider a few nonlinear system identification problems. The first example is on the parameter cum state estimation of a hardening Duffing oscillator modeled under additive stochastic excitation with constant coefficients. The next problem involves parameter/state estimation of yet another mechanical oscillator, a higher dimensional shear frame under harmonic dynamic loading. The third problem is on target tracking with large and even non-Gaussian observation noise. In all the cases, the performance of the KS filter has been assessed through comparisons with a few competing schemes, e.g. the auxiliary bootstrap filter (ABS) [30] and the ensemble Kalman filter (EnKF) [31].

### 3.1 Example 1: A Hardening Duffing Oscillator

The system dynamics of a suitably parameterized Duffing oscillator [32] under additive noise is given by:

$$\ddot{x} + 2\pi\varepsilon_1 \dot{x} + 4\pi^2\varepsilon_2\left(1+x^2\right)x = 4\pi^2\varepsilon_3 \cos(2\pi t) + f\,\dot{B}(t) \tag{3.1}$$

The state space representation of the above equation in incremental form yields the process SDE as:

$$\begin{aligned}dx_1(t) &= x_2(t)\,dt \\ dx_2(t) &= \left(-cx_2 - kx_1 - \alpha x_1^3 + 4\pi^2\varepsilon_3 \cos(2\pi t)\right)dt + f\,dB(t)\end{aligned} \tag{3.2}$$

where $x_1 := x, x_2 := \dot{x},\ c := 2\pi\varepsilon_1, k = \alpha := 4\pi^2\varepsilon_2$ (even though their reference values are the same, $k$ and $\alpha$ are treated as two separate parameters for estimation purposes). The aim is to estimate the parameter vector $\mu := \{k\ \ c\ \ \alpha\}^T$ as well as the displacement and velocity states given only the observed displacement history. The associated observation SDE is written as:

$$dY_t = \sigma_m^{-1} x_1 dt + dW_t \tag{3.3}$$

where $W_t$ is a standard Brownian motion and $Y_t$, adopted as the measured entity for the filter, is computable as:

$$Y_t = \int_0^t \sigma_m^{-1} y_s ds \tag{3.4}$$

Evaluation of $Y_t$ as above needs a suitable numerical integration scheme. Here $y_t$ denotes the actually observed displacement at discrete time instants and is synthetically generated by corrupting $x_1(t)$ with observation noise, obtainable through a higher order integration scheme [33] applied to the process SDE. Reference values of $\varepsilon_1$, $\varepsilon_2$ and $\varepsilon_3$ used to generate $y_t$ are 0.25, 1.0, 5.0 respectively. For parameter estimation, the process model (3.2) is augmented with the following additional set of SDEs:

$$d\mu = f_\mu dB_\mu \tag{3.5}$$

where $B_\mu \in \mathbb{R}^3$ is a standard Brownian motion used to model the noise process associated with the parameter evolution and $f_\mu$ a $3 \times 3$ diagonal matrix of constant entries, representing the noise intensities. The process state, so augmented, is denoted by $X_t = \{x_1 \ x_2 \ k \ c \ \alpha\}^T$, and we aim at finding the estimate:

$$\pi_t(X_t) := E_P(X_t | \mathcal{F}_t^Y)$$

The KS filter is recursively implemented following the prediction and iterated updating stages described in Section 2 and the results are reported below.

Figs 3.1(a)-3.1(c) show the reconstruction of the parameters $(k, c, \alpha)$ via the KS filter and the ABS. For the ABS, the filtering problem is solved with two different observation models, namely ABS1 and ABS2. For ABS2, the model is the same as the one used with the KS filter (i.e. the SDE given by Eq. 3.3). In ABS1, on the other hand, an algebraic observation model, which avoids the integration error introduced in the SDE-based model of Eq. (3.3), is made use of. The discrete-time form of the last model is given by:

$$y(t_i) = x_1(t_i) + \sigma_m \eta_i \tag{3.6}$$

where $\eta_i \sim \mathcal{N}(0,1)$ is a standard normal random variable. The reported parameter estimates in Fig. 3.1 clearly reveal the substantively improved convergence and accuracy of the KS filter over the twin ABS schemes. Specifically, comparing the results of KS and ABS2, which work with the same observation model, the robustness of the KS filter may be inferred from the fact that its performance is only marginally affected by the integration error in evaluating $Y_t$ (in Eq. 3.4). Despite the observation integration error, a comparison of the results of the KS and ABS1 filters continues to confirm the superior features of the former. This is further verified in Figs 3.2(a)-3.2(e), wherein the sampling standard deviation plots for the estimates are reported. For each filter run, 200 Monte Carlo particles are used and 100 different such filter runs are utilized to obtain the (sampling) standard deviation plots.

It is further noted that the KS filter has been implemented with both the schedules for $\{\beta_k\}$ as described in Section 2. While the expensive annealing schedule of [27] performs better than that based on exponential decay, the reconstruction results with the latter are only marginally affected owing to an ensemble of Markov chains, which by themselves provide a reasonable exploration of the phase space. Since the faster decay in $\{\beta_k\}$ prescribed by the exponential scheme appears to insignificantly impact the performance of the KS filter, only 10 inner iterations are consistently used through the numerical work reported in this section.

### 3.2 Example 2: A Five story Shear Frame under Dynamic Loading

For the state-parameter estimation of a 5-story shear frame, the system model is considered to be of the form:

$$\ddot{X}(t)+[C]\dot{X}(t)+[K]X(t)+\tilde{X}_{nl} = F(t)+\sigma \dot{B} \tag{3.6}$$

The stiffness matrix is given as $[K] = \begin{bmatrix} K_1+K_2 & -K_2 & 0 & 0 & 0 \\ -K_2 & K_2+K_3 & -K_3 & 0 & 0 \\ 0 & -K_3 & K_3+K_4 & -K_4 & 0 \\ 0 & 0 & -K_4 & K_4+K_5 & -K_5 \\ 0 & 0 & 0 & -K_5 & K_5 \end{bmatrix}$ and the viscous damping matrix $[C]$ is obtained by replacing $K_i$ by $C_i$ in the above matrix, where $K_i$ and $C_i$ are respectively the stiffness and damping parameters corresponding to the

$i^{th}$ floor of the frame. $F(t) \in \mathbb{R}^5$ is a deterministic forcing vector and $\tilde{X}_{nl}$ is a known nonlinear field (containing polynomial terms in the elements of $X$) introduced in the model. The aim is to estimate the stiffness and damping coefficients as well as the velocity and displacement vectors, conditioned on only the measured displacements of the floors. For purposes of comparison, only the ABS1 filter is used as numerical simulations via the ABS2 are often found to either diverge or quickly degenerate to a single particle location owing to weight collapse for this 20-dimensional filtering problem.

Figs 3.3(a)-3.3(d) show a comparison of the estimated states and parameters corresponding to the $4^{th}$ floor of the frame via the KS and ABS1 filters. The results reveal that while the ABS1 does provide non-trivial, albeit erroneous, updates to the displacement/velocity states, it does suffer from a partial degeneracy, which is evidenced through its failure to provide non-zero updates to the parameters. Even a 10-fold increase in the ensemble size (from 200 to 2000) fails to arrest the weight collapse. This may be well contrasted with the KS filter, which successfully estimates all the states and parameters even with a moderate ensemble size of $N = 200$. Although results corresponding to the $4^{th}$ floor are only reported, a similar trend is observed for all the system/parameter states. Accuracy and convergence apart, the KS filter also demonstrates a non-trivially reduced sampling variance in the estimates and this is brought forth via sampling standard deviation plots for the estimates in Figs 3.4(a) – 3.4(d). Here, for each filter run, 200 particles are used and 100 independent filter runs are utilized to obtain the standard deviation plots.

### *3.3 Example 3: A Target Tracking Problem*

In this problem, we estimate the trajectory of a target (i.e. its position and velocity) from the highly noisy sensor data. The target is assumed to move in the x-y plane with velocity $[\dot{x} \quad \dot{y}]^T$ and its dynamics is given by:

$$X_{i+1} = FX_i + \Gamma a_i$$

where $X_i = [x \ \dot{x} \ y \ \dot{y}]^T_i$, $F = \begin{bmatrix} 1 & \Delta T & 0 & 0 \\ 0 & 1 & 0 & 0 \\ 0 & 0 & 1 & \Delta T \\ 0 & 0 & 0 & 1 \end{bmatrix}$ and $\Gamma = \begin{bmatrix} \frac{1}{2}\Delta T^2 & 0 \\ 1 & 0 \\ 0 & \frac{1}{2}\Delta T^2 \\ 0 & 1 \end{bmatrix}$. $a_i$ is the random acceleration of the target and, in the present problem, is characterized by a white noise. The distance and angle from the observer situated at the origin, $(x_0, y_0)$ are taken as the measurements and the corresponding observation equation is given, in the algebraic form, by

$$Z_i = \begin{bmatrix} \tan^{-1}\left(\frac{y_i - y_0}{x_i - x_0}\right) \\ \sqrt{(y_i - y_0)^2 + (x_i - x_0)^2} \end{bmatrix} + v_i$$

where $v_i \sim \mathcal{N}(0, S_v)$ is the measurement noise. $S_v$ is the noise intensity taken as 5% of the measurement. The target starts at location [0.5 1] (m) in Cartesian coordinates with initial velocity [3 1] (in m/s) and experiences a 4-leg maneuvering sequence by taking sharp turns at 20sec ([-40 40] m/s$^2$), 30 sec([25 -25] m/s$^2$), 60 sec([25 -25] m/s$^2$) and 80 sec([-30 30] m/s$^2$) respectively. Considering the target as a rigid body, the discrete process equation is here given by the simple motion model $X_{i+1} = FX_i + \Gamma w_i$, where $w_i$ is a white noise with intensity 1. The performance of the KS filter (again with $N = 200$), is assessed against both an ensemble Kalman filter (with 200 particles) and the auxiliary bootstrap filter (the ABS1 version; using 200 and 2000 particles). Curiously, the EnKF, like the KS filter, is a gain-based filter that also bypasses the problem of weight collapse. Tracking estimates reported in Fig 3.5 confirms that the KS filter not only outperforms the ABS filter, but also the EnKF.

The reduced variance in the estimates of the KS filter, borne out by the above examples, is suggestive of its ability to handle estimation problems with fewer particles. It is therefore interesting to further investigate on the minimum ensemble size that could yield acceptable results. Specifically, it may be observed that a one-particle simulation (i.e. $N = 1$) is infeasible as it would identically yield a zero gain (in Eq. 2.6). In figure 3.6, with $N = 5$, the performance of the KS filter is compared with the EnKF in tracking the same trajectory as above but with a (non-Gaussian) glint noise. The process and measurement equations are as given above, except that the glint measurement noise $v_i$ is modeled by a Gaussian mixture as $p(v_i) = (1-\gamma)\mathcal{N}(0, S_v) + \gamma \mathcal{N}(0, S_v)$, with a glint noise probability $\gamma = 0.5$. Although we have used the exponential decay of the ATP in this example, the reduced ensemble size might necessitate the use of a rigorous annealing schedule (as prescribed in [27]) in other problems.

A consistent observation across our numerical work has been that the KS filter works quite well across a broad range of measurement noise intensity – from high to very low. Filtering solutions corresponding to low measurement noise intensity are of particular interest, as most particle filters fail to work satisfactorily in this regime owing to an accelerated development of the weight degeneracy problem. Moreover, relatively reduced sampling fluctuations in the estimates via the KS filter may be attributed to the additive nature of the nonlinear update, which not only provide freedom from the weight degeneracy of most particle filters but also from the limitations of the EnKF-type additive updates, derived essentially from the Kalman filter and providing accurate solutions only for linear measurement models with Gaussian diffusion terms.

## 4 Conclusions

As with most other particle filters, the KS filter may be viewed as being based on a particle form of the Kushner-Stratonovich equation, which governs the evolution of estimates (or the filtering density) for a nonlinear, non-Gaussian filtering problem. The first point of departure in the present proposal is however in exploiting a suitably derived discrete form of the gain-like integral in the KS equation to provide non-Newton (i.e. derivative-free) directional information towards additively updating the predicted solution. The second, and by far the more important, feature of the KS filter is inner updating method used to drive the time-discrete innovation, over every time step, to a zero-mean Brownian increment whilst employing an artificially introduced time-like parameter. The annealing-type schedule of the scalar diffusion parameter may however be chosen far more flexibly than conventional annealing schedules characterized by a slower decay of the parameter. Detailed error estimates for the approximations corresponding to time discretization, inner iteration and ensemble finiteness are provided. As demonstrated through numerical explorations on a few nonlinear dynamical system identification problems, these novel features engender improved convergence characteristics in the KS filter, conjoined with higher estimation accuracy and reduced sampling variance of the estimates. The limited numerical evidence from this work is also suggestive of the potential applicability of the KS filtering scheme to a broad array of higher dimensional system identification problems of practical engineering interest.

**Acknowledgements:** The authors wish to acknowledge the role played by the constructive comments of all the three anonymous reviewers in substantively improving the quality of the presented material.

## Appendix I

**Proof of Theorem I**

All constants used in the derivation are independent of $\Delta t$ and $N$. The a-priori error is estimated using Minkowski's inequality,

$$\left(E_P\left[\left|\pi_{i+1}(\phi)-\pi_{i+1,\kappa}^{\prime e}(\phi)\right|^2\right]\right)^{\frac{1}{2}} \leq \left(E_P\left[\left|\pi_{i+1}(\phi)-\pi_{i+1,1}^{\prime e}(\phi)\right|^2\right]\right)^{\frac{1}{2}} + \left(E_P\left[\left|\pi_{i+1,1}^{\prime e}(\phi)-\pi_{i+1,\kappa}^{\prime e}(\phi)\right|^2\right]\right)^{\frac{1}{2}} \quad \text{(A1)}$$

The first term $\left(E_P\left[\left|\pi_{i+1}(\phi)-\pi_{i+1,1}^{\prime e}(\phi)\right|^2\right]\right)^{\frac{1}{2}}$, on the right hand side of A1, corresponds to the error in prediction along with the initial update, based on the EM approximation of the KS equation within the MC framework. This term may further be split into two parts, one corresponding to EM integration and the other due to ensemble approximation:

$$\left(E_P\left[\left|\pi_{i+1}(\phi)-\pi_{i+1,1}^{\prime e}(\phi)\right|^2\right]\right)^{\frac{1}{2}} \leq \left(E_P\left[\left|\pi_{i+1}(\phi)-\pi_{i+1,1}^{e}(\phi)\right|^2\right]\right)^{\frac{1}{2}} + \left(E_P\left[\left|\pi_{i+1,1}^{e}(\phi)-\pi_{i+1,1}^{\prime e}(\phi)\right|^2\right]\right)^{\frac{1}{2}} \quad \text{(A2)}$$

We first obtain an error bound for the time discretization.

*Lemma 1:*

If the process has bounded moments of any order and $\phi \in C_b^2(\mathbb{R})$, then

$$\left(E_P\left[\left|\pi_{i+1}(\phi)-\pi_{i+1,1}^{e}(\phi)\right|^2\right]\right)^{\frac{1}{2}} \leq D'(\Delta t)^{\frac{1}{2}}$$

*Proof:*

Using the conditional version of Jensen's inequality, we have

$$E_P\left[\left|\pi_{i+1}(\phi) - \pi^e_{i+1,1}(\phi)\right|^2\right] = E_P\left[\left|E_P\left[\phi_{i+1} - \phi^e_{i+1,1}\right] | \mathcal{F}^Y_{t_{i+1}}\right|^2\right]$$

$$\leq E_P\left[E_P\left[\left|\phi_{i+1} - \phi^e_{i+1,1}\right|^2 | \mathcal{F}^Y_{t_{i+1}}\right]\right]$$

$$= E_P\left[\left|\phi_{i+1} - \phi^e_{i+1,1}\right|^2\right]$$

Using the standard strong order of convergence of the EM method [34], we obtain

$$\left(E_P\left[\left|\hat{X}_{i+1} - \hat{X}^e_{i+1,1}\right|^2\right]\right)^{\frac{1}{2}} \leq D_1 (\Delta t)^{\frac{1}{2}} \tag{A3}$$

where $D_1 > 0$ is a constant independent of $\Delta t$. In general, for $p \geq 1, D_2 > 0$, one can write:

$$E_P\left[\left|\hat{X}_{i+1} - \hat{X}^e_{i+1,1}\right|^{2p}\right]^{\frac{1}{2p}} \leq D_2 (\Delta t)^{\frac{1}{2}} \tag{A4}$$

Furthermore, we assume that $\phi$ is sufficiently smooth so that $\phi(x)$ and its derivatives satisfy an inequality of the form

$$|\phi(x)| \leq M_3 \left(1 + |x|^a\right) \tag{A5}$$

for positive constants $M_3, a$. Hence we can write

$$\left|\phi_{i+1} - \phi^e_{i+1,1}\right| \leq D_3 \left(1 + \left|\hat{X}_{i+1}\right|^a + \left|\hat{X}^e_{i+1,1}\right|^a\right)\left|\hat{X}_{i+1} - \hat{X}^e_{i+1,1}\right| \tag{A6}$$

where $D_3 > 0$. Then, using the Cauchy-Schwarz inequality, we have

$$E_P\left[\left|\phi_{i+1} - \phi^e_{i+1,1}\right|^{2p}\right] \leq (D_3)^{2p} E_P\left[\left(1 + \left|\hat{X}_{i+1}\right|^a + \left|\hat{X}^e_{i+1,1}\right|^a\right)^{2p} \left|\hat{X}_{i+1} - \hat{X}^e_{i+1,1}\right|^{2p}\right]$$

$$\leq (D_3)^{2p} \sqrt{E_P\left[\left(1 + \left|\hat{X}_{i+1}\right|^a + \left|\hat{X}^e_{i+1,1}\right|^a\right)^{4p}\right]} \sqrt{E_P\left[\left|\hat{X}_{i+1} - \hat{X}^e_{i+1,1}\right|^{4p}\right]} \quad (A7)$$

$$\leq D_4\left(\hat{X}_{i+1}\right)\left((\Delta t)^{\frac{1}{2}}\right)^{2p}$$

Hence, for $p = 1$, we get $\left(E_P\left[\left|\pi_{i+1}(\phi) - \pi^e_{i+1,1}(\phi)\right|^2\right]\right)^{\frac{1}{2}} \leq D'(\Delta t)^{\frac{1}{2}}.$ (A8)

□

Next, we consider the error due to the ensemble approximation within the time-discrete framework. In the recursive setting, given the empirical filtered distribution of $X_t$ at $t = t_i$, we may consider:

$$E_P\left[\left|\pi^e_i(\phi) - \pi'^e_i(\phi)\right|^2\right] \leq D_5 \frac{\|\phi\|^2}{N}$$

Recall that $\|\cdot\|$ denotes the supremum norm on $C_b(\mathbb{R})$.

*Lemma 2:*

Assume that for any $\phi \in C_b^2(\mathbb{R})$,

$$E_P\left[\left|\pi^e_i(\phi) - \pi'^e_i(\phi)\right|^2\right] \leq D_5 \frac{\|\phi\|^2}{N}$$

Then,

$$\left(E_P\left[\left|\pi'^e_{t_{i+1},1}(\phi) - \pi^e_{t_{i+1},1}(\phi)\right|^2\right]\right)^{\frac{1}{2}} \leq \frac{D_6}{\sqrt{N}} \begin{pmatrix} (3|\Delta Y_{i+1}| + 5\|h\|)\|h\|\|\phi\| + (|\Delta Y_{i+1}|\Delta t)\|L(h\phi)\| \\ + (|\Delta Y_{i+1}|\Delta t\|h\| + (\Delta t)^2\|h\|^2)\|L(\phi)\| \\ + (|\Delta Y_{i+1}|\Delta t\|\phi\| + 4(\Delta t)^2\|h\|\|\phi\|)\|L(h)\| + \Delta t\|L(\phi)\| + \|\phi\| \end{pmatrix}$$

*Proof:*

Using Minkowski's inequality, we can write,

$$\left(E_P\left[\left|\pi_{i+1,1}^{\prime e}(\phi)-\pi_{i+1,1}^{e}(\phi)\right|^{2}\right]\right)^{\frac{1}{2}} \leq \left(E_P\left[\left|\{\tilde{\pi}_{i+1}^{\prime e}(\phi)+G_{i+1}^{\prime}(\phi)\}-\{\tilde{\pi}_{i+1}^{\prime e}(\phi)+G_{i+1}(\phi)\}\right|^{2}\right]\right)^{\frac{1}{2}}$$

$$+\left(E_P\left[\left|\{\tilde{\pi}_{i+1}^{\prime e}(\phi)+G_{i+1}(\phi)\}-\{\tilde{\pi}_{i+1}^{e}(\phi)+G_{i+1}(\phi)\}\right|^{2}\right]\right)^{\frac{1}{2}}$$
(A9)

where $G_{i+1}^{\prime}(\phi) = \{\tilde{\pi}_{i+1}^{\prime e}(h\phi) - \tilde{\pi}_{i+1}^{\prime e}(h)\tilde{\pi}_{i+1}^{\prime e}(\phi)\}\{\Delta Y_i - \tilde{\pi}_{i+1}^{\prime e}(h)\Delta t\}$ and similarly one can write the expression for $G_{i+1}(\phi)$ by appropriately replacing the ensemble approximation in $G_{i+1}^{\prime}(\phi)$. $\tilde{\pi}_{i+1}^{e}(\phi)$ is the predicted conditional estimate at $t_{i+1}$. Its ensemble approximation is denoted as $\tilde{\pi}_{i+1}^{\prime e}(\phi)$. Now, again using Minkowski's inequality, we have

$$\left(E_P\left[\left|\tilde{\pi}_{i+1}^{\prime e}(\phi) - \tilde{\pi}_{i+1}^{e}(\phi)\right|^{2}\right]\right)^{\frac{1}{2}} \leq \left(E_P\left[\left|\tilde{\pi}_{i+1}^{\prime e}(\phi) - \left(\pi_i^{\prime e}(\phi) + \pi_i^{e}(\mathrm{L}(\phi))\Delta t\right)\right|^{2}\right]\right)^{\frac{1}{2}}$$

$$+\left(E_P\left[\left|\left(\pi_i^{\prime e}(\phi) + \pi_i^{e}(\mathrm{L}(\phi))\Delta t\right) - \tilde{\pi}_{i+1}^{e}(\phi)\right|^{2}\right]\right)^{\frac{1}{2}}$$

$$\leq \Delta t\left(E_P\left[\left|\pi_i^{\prime e}(\mathrm{L}(\phi)) - \pi_i^{e}(\mathrm{L}(\phi))\right|^{2}\right]\right)^{\frac{1}{2}} + \left(E_P\left[\left|\pi_i^{\prime e}(\phi) - \pi_i^{e}(\phi)\right|^{2}\right]\right)^{\frac{1}{2}}$$

$$\leq D_6(\Delta t\frac{\|\mathrm{L}(\phi)\|}{\sqrt{N}} + \frac{\|\phi\|}{\sqrt{N}})$$
(A10)

where the generator L has been defined in Eqn. 2.1c. Having obtained a bound for the ensemble approximated prediction, we now proceed to get a similar error bound for the update using A9.

$$\left(E_P\left[\left|G_{t_{i+1}}^{\prime}(\phi) - G_{t_{i+1}}(\phi)\right|^{2}\right]\right)^{\frac{1}{2}} = \left(E_P\left[\left|\begin{array}{l}\{\tilde{\pi}_{i+1}^{\prime e}(h\phi) - \tilde{\pi}_{i+1}^{\prime e}(h)\tilde{\pi}_{i+1}^{\prime e}(\phi)\}\{\Delta Y_{i+1} - \tilde{\pi}_{i+1}^{\prime e}(h)\Delta t\} - \\ \{\tilde{\pi}_{i+1}^{e}(h\phi) - \tilde{\pi}_{i+1}^{e}(h)\tilde{\pi}_{i+1}^{e}(\phi)\}\{\Delta Y_{i+1} - \tilde{\pi}_{i+1}^{e}(h)\Delta t\}\end{array}\right|^{2}\right]\right)^{\frac{1}{2}}$$

$$\leq \left|\Delta Y_{i+1}\right|\left(E_P\left[\left|\{\tilde{\pi}_{i+1}^{\prime e}(h\phi) - \tilde{\pi}_{i+1}^{\prime e}(h)\tilde{\pi}_{i+1}^{\prime e}(\phi)\} - \{\tilde{\pi}_{i+1}^{e}(h\phi) - \tilde{\pi}_{i+1}^{e}(h)\tilde{\pi}_{i+1}^{e}(\phi)\}\right|^{2}\right]\right)^{\frac{1}{2}}$$

$$+\Delta t\left(E_P\left[\left|\{\tilde{\pi}_{i+1}^{\prime e}(h\phi) - \tilde{\pi}_{i+1}^{\prime e}(h)\tilde{\pi}_{i+1}^{\prime e}(\phi)\}\tilde{\pi}_{i+1}^{\prime e}(h) - \{\tilde{\pi}_{i+1}^{e}(h\phi) - \tilde{\pi}_{i+1}^{e}(h)\tilde{\pi}_{i+1}^{e}(\phi)\}\tilde{\pi}_{i+1}^{e}(h)\right|^{2}\right]\right)^{\frac{1}{2}}$$

(A11)

Here,

$$\left(E_P\left[\left|\left\{\tilde{\pi}_{i+1}^{\prime e}(h\phi) - \tilde{\pi}_{i+1}^{\prime e}(h)\tilde{\pi}_{i+1}^{\prime e}(\phi)\right\} - \left\{\tilde{\pi}_{i+1}^{e}(h\phi) - \tilde{\pi}_{i+1}^{e}(h)\tilde{\pi}_{i+1}^{e}(\phi)\right\}\right|^2\right]\right)^{\frac{1}{2}}$$

$$\leq \left(E_P\left[\left|\left\{\tilde{\pi}_{i+1}^{\prime e}(h\phi) - \tilde{\pi}_{i+1}^{e}(h\phi)\right\}\right|^2\right]\right)^{\frac{1}{2}} + \left(E_P\left[\left|\left\{\tilde{\pi}_{i+1}^{e}(h)\tilde{\pi}_{i+1}^{e}(\phi) - \tilde{\pi}_{i+1}^{\prime e}(h)\tilde{\pi}_{i+1}^{\prime e}(\phi)\right\}\right|^2\right]\right)^{\frac{1}{2}}$$

(A12)

However, using A10, we may write

$$\left(E_P\left[\left|\left\{\tilde{\pi}_{i+1}^{\prime e}(h\phi) - \tilde{\pi}_{i+1}^{e}(h\phi)\right\}\right|^2\right]\right)^{\frac{1}{2}} \leq D_6(\Delta t \frac{\|L(h\phi)\|}{\sqrt{N}} + \frac{\|h\phi\|}{\sqrt{N}}) \quad (A13)$$

Now focusing on the second term on the right hand side of A12, we have

$$E_P\left[\left|\left\{\tilde{\pi}_{i+1}^{\prime e}(h)\tilde{\pi}_{i+1}^{\prime e}(\phi) - \tilde{\pi}_{i+1}^{e}(h)\tilde{\pi}_{i+1}^{e}(\phi)\right\}\right|^2\right]^{\frac{1}{2}} \leq E_P\left[\left|\left\{\tilde{\pi}_{i+1}^{\prime e}(h)\tilde{\pi}_{i+1}^{\prime e}(\phi) - \tilde{\pi}_{i+1}^{\prime e}(h)\tilde{\pi}_{i+1}^{e}(\phi)\right\}\right|^2\right]^{\frac{1}{2}}$$

$$+ E_P\left[\left|\left\{\tilde{\pi}_{i+1}^{\prime e}(h)\tilde{\pi}_{i+1}^{e}(\phi) - \tilde{\pi}_{i+1}^{e}(h)\tilde{\pi}_{i+1}^{e}(\phi)\right\}\right|^2\right]^{\frac{1}{2}}$$

$$\leq \|h\| E_P\left[\left|\left\{\tilde{\pi}_{i+1}^{\prime e}(\phi) - \tilde{\pi}_{i+1}^{e}(\phi)\right\}\right|^2\right]^{\frac{1}{2}} + \|\phi\| E_P\left[\left|\left\{\tilde{\pi}_{i+1}^{\prime e}(h) - \tilde{\pi}_{i+1}^{e}(h)\right\}\right|^2\right]^{\frac{1}{2}}$$

$$\leq D_6 \left(\Delta t \frac{\|h\|\|L(\phi)\|}{\sqrt{N}} + \frac{\|h\|\|\phi\|}{\sqrt{N}} + \Delta t \frac{\|\phi\|\|L(h)\|}{\sqrt{N}} + \frac{\|\phi\|\|h\|}{\sqrt{N}}\right)$$

(A14)

Now consider the second term on the right hand side of A11:

$$\left(E_P\left[\left|\left\{\tilde{\pi}_{i+1}^{\prime e}(h\phi) - \tilde{\pi}_{i+1}^{\prime e}(h)\tilde{\pi}_{i+1}^{\prime e}(\phi)\right\}\tilde{\pi}_{i+1}^{\prime e}(h) - \left\{\tilde{\pi}_{i+1}^{e}(h\phi) - \tilde{\pi}_{i+1}^{e}(h)\tilde{\pi}_{i+1}^{e}(\phi)\right\}\tilde{\pi}_{i+1}^{e}(h)\right|^2\right]\right)^{\frac{1}{2}}$$

$$\leq \left(E_P\left[\left|\left\{\tilde{\pi}_{i+1}^{\prime e}(h\phi)\tilde{\pi}_{i+1}^{\prime e}(h) - \tilde{\pi}_{i+1}^{e}(h\phi)\tilde{\pi}_{i+1}^{e}(h)\right\}\right|^2\right]\right)^{\frac{1}{2}} + \left(E_P\left[\left|\left\{\left(\tilde{\pi}_{i+1}^{\prime e}(h)\right)^2 \tilde{\pi}_{i+1}^{\prime e}(\phi) - \left(\tilde{\pi}_{i+1}^{e}(h)\right)^2 \tilde{\pi}_{i+1}^{e}(\phi)\right\}\right|^2\right]\right)^{\frac{1}{2}}$$

(A15)

Deriving in the same way as in A14, we get for the first term on the right hand side

$$\left(E_P\left[\left|\left\{\tilde{\pi}_{i+1}^{\prime e}(h\phi)\tilde{\pi}_{i+1}^{\prime e}(h)-\tilde{\pi}_{i+1}^{e}(h\phi)\tilde{\pi}_{i+1}^{e}(h)\right\}\right|^2\right]\right)^{\frac{1}{2}} \leq D_6\left(\Delta t\frac{\|h\phi\|\|\mathrm{L}(h)\|}{\sqrt{N}}+\frac{\|h\phi\|\|h\|}{\sqrt{N}}+\Delta t\frac{\|h\phi\|\|\mathrm{L}(h)\|}{\sqrt{N}}+\frac{\|h\phi\|\|h\|}{\sqrt{N}}\right)$$

(A16)

For the second term on the right hand side of A15, we have

$$\left(E_P\left[\left|\left\{\left(\tilde{\pi}_{i+1}^{\prime e}(h)\right)^2\tilde{\pi}_{i+1}^{\prime e}(\phi)-\left(\tilde{\pi}_{i+1}^{e}(h)\right)^2\tilde{\pi}_{i+1}^{e}(\phi)\right\}\right|^2\right]\right)^{\frac{1}{2}}$$

$$\leq \left(E_P\left[\left|\left\{\left(\tilde{\pi}_{i+1}^{\prime e}(h)\right)^2\tilde{\pi}_{i+1}^{\prime e}(\phi)-\left(\tilde{\pi}_{i+1}^{e}(h)\right)^2\tilde{\pi}_{i+1}^{\prime e}(\phi)\right\}\right|^2\right]\right)^{\frac{1}{2}}+\left(E_P\left[\left|\left\{\left(\tilde{\pi}_{i+1}^{e}(h)\right)^2\tilde{\pi}_{i+1}^{\prime e}(\phi)-\left(\tilde{\pi}_{i+1}^{e}(h)\right)^2\tilde{\pi}_{i+1}^{e}(\phi)\right\}\right|^2\right]\right)^{\frac{1}{2}}$$

$$\leq \|\phi\|\left(E_P\left[\left|\left\{\left(\tilde{\pi}_{i+1}^{\prime e}(h)\right)^2-\left(\tilde{\pi}_{i+1}^{e}(h)\right)^2\right\}\right|^2\right]\right)^{\frac{1}{2}}+\|h\|^2\left(E_P\left[\left|\left\{\tilde{\pi}_{i+1}^{\prime e}(\phi)-\tilde{\pi}_{i+1}^{e}(\phi)\right\}\right|^2\right]\right)^{\frac{1}{2}}$$

$$\leq D_6(2\Delta t\frac{\|h\|\|\phi\|\|\mathrm{L}(h)\|}{\sqrt{N}}+2\frac{\|\phi\|\|h\|^2}{\sqrt{N}}+\Delta t\frac{\|h\|^2\|\mathrm{L}(\phi)\|}{\sqrt{N}}+\frac{\|h\|^2\|\phi\|}{\sqrt{N}})$$

(A17)

Putting the above error bounds together in A11, one obtains

$$\left(E_P\left[\left|G'_{t_{i+1}}(\phi)-G_{t_{i+1}}(\phi)\right|^2\right]\right)^{\frac{1}{2}} \leq D_6\,|\Delta Y_{i+1}|\begin{pmatrix}\Delta t\dfrac{\|\mathrm{L}(h\phi)\|}{\sqrt{N}}+\dfrac{\|h\phi\|}{\sqrt{N}}+\\ \left(\Delta t\dfrac{\|h\|\|\mathrm{L}(\phi)\|}{\sqrt{N}}+\dfrac{\|h\|\|\phi\|}{\sqrt{N}}+\Delta t\dfrac{\|\phi\|\|\mathrm{L}(h)\|}{\sqrt{N}}+\dfrac{\|\phi\|\|h\|}{\sqrt{N}}\right)\end{pmatrix}$$

$$+D_6\Delta t\begin{pmatrix}\left(\Delta t\dfrac{\|h\phi\|\|\mathrm{L}(h)\|}{\sqrt{N}}+\dfrac{\|h\phi\|\|h\|}{\sqrt{N}}+\Delta t\dfrac{\|h\phi\|\|\mathrm{L}(h)\|}{\sqrt{N}}+\dfrac{\|h\phi\|\|h\|}{\sqrt{N}}\right)\\ +2\Delta t\dfrac{\|h\|\|\phi\|\|\mathrm{L}(h)\|}{\sqrt{N}}+2\dfrac{\|\phi\|\|h\|^2}{\sqrt{N}}+\Delta t\dfrac{\|h\|^2\|\mathrm{L}(\phi)\|}{\sqrt{N}}+\dfrac{\|h\|^2\|\phi\|}{\sqrt{N}}\end{pmatrix}$$

(A18)

On simplifying the terms, we get:

$$\left(E_P\left[\left|G'_{t_{i+1}}(\phi)-G_{t_{i+1}}(\phi)\right|^2\right]\right)^{\frac{1}{2}} \leq \frac{D_6}{\sqrt{N}}\left(\begin{array}{l}|\Delta Y_{i+1}|\left(3\|h\|\|\phi\|+\Delta t\|L(h\phi)\|+\Delta t\|h\|\|L(\phi)\|+\Delta t\|\phi\|\|L(h)\|\right)\\+\left(5\|h\|^2\|\phi\|\Delta t+4(\Delta t)^2\|h\|\|\phi\|\|L(h)\|+(\Delta t)^2\|h\|^2\|L(\phi)\|\right)\end{array}\right)$$

$$\leq \frac{D_6}{\sqrt{N}}\left(\begin{array}{l}(3|\Delta Y_{i+1}|+5\|h\|)\|h\|\|\phi\|+(|\Delta Y_{i+1}|\Delta t)\|L(h\phi)\|+\left(|\Delta Y_{i+1}|\Delta t\|h\|+(\Delta t)^2\|h\|^2\right)\|L(\phi)\|\\+\left(|\Delta Y_{i+1}|\Delta t\|\phi\|+4(\Delta t)^2\|h\|\|\phi\|\right)\|L(h)\|\end{array}\right)$$

(A19)

Thus we have the following bound for the ensemble approximation error:

$$\left(E_P\left[\left|\pi'^e_{t_{i+1},1}(\phi)-\pi^e_{t_{i+1},1}(\phi)\right|^2\right]\right)^{\frac{1}{2}} \leq \frac{D_6}{\sqrt{N}}\left(\begin{array}{l}(3|\Delta Y_{i+1}|+5\|h\|)\|h\|\|\phi\|+(|\Delta Y_{i+1}|\Delta t)\|L(h\phi)\|\\+\left(|\Delta Y_{i+1}|\Delta t\|h\|+(\Delta t)^2\|h\|^2\right)\|L(\phi)\|\\+\left(|\Delta Y_{i+1}|\Delta t\|\phi\|+4(\Delta t)^2\|h\|\|\phi\|\right)\|L(h)\|+\Delta t\|L(\phi)\|+\|\phi\|\end{array}\right) \quad (A20)$$

Hence, from A2 we get

$$\left(E_P\left[\left|\pi_{t_{i+1}}(\phi)-\pi'^e_{t_{i+1},1}(\phi)\right|^2\right]\right)^{\frac{1}{2}} \leq D'(\Delta t)^{\frac{1}{2}}$$

$$+\frac{D''}{\sqrt{N}}\left(\begin{array}{l}(3|\Delta Y_{i+1}|+5\|h\|)\|h\|\|\phi\|+(|\Delta Y_{i+1}|\Delta t)\|L(h\phi)\|\\+\left(|\Delta Y_{i+1}|\Delta t\|h\|+(\Delta t)^2\|h\|^2\right)\|L(\phi)\|\\+\left(|\Delta Y_{i+1}|\Delta t\|\phi\|+4(\Delta t)^2\|h\|\|\phi\|\right)\|L(h)\|+\Delta t\|L(\phi)\|+\|\phi\|\end{array}\right) \quad (A21)$$

where $D''=D_6$. □

The second error term in the right hand side of the inequality A1 may be bounded by suitably characterizing the inner iterations (chains). These chains are propagated in such a way that, upon convergence, they lead to the filtered distribution at a given time $t_{i+1}$. In the following derivation, definitions 1, 2, proposition 1 and theorem 2 are fashioned after [36].

*Convergence of the inner iterations*

One way to prove that the inner iterations converge would be to treat them as stochastic Picard iterations, wherein the aim would be to demonstrate that the sequence so generated is Cauchy and hence the associated map has a fixed point that corresponds to a solution of the KS equation. However, in this work, we adopt a different route, wherein the effort is to characterize the transitional measures associated with the sequences (or chains) generated by the iterations (over $\vartheta$) and thus to establish that the targeted filtered distribution over a given time step may be arrived at as a stationary distribution.

*Definition 1*:

Inner iteration on the sample space $\Omega = \mathbb{R}$ (or on the Borel $\sigma$-algebra $\mathcal{F}$) is reversible with respect to a probability distribution $\pi_{i+1}(\cdot)$ on, if

$$\pi_{i+1}(dx)\mathbf{P}_{i+1}(x,dy) = \pi_{i+1}(dy)\mathbf{P}_{i+1}(x,dy); \quad x, y \in \mathbb{R} \tag{A22}$$

Here $\mathbf{P}_{i+1}(x,dy)$ denotes a suitably defined transitional measure. First we need to show that the inner iteration at a given time instant $t_{i+1}$ corresponds to the stationary distribution $\pi_{i+1}(\cdot)$.

*Proposition 1*: [Proposition 1 of [36]]

If the inner iteration is reversible with respect to $\pi_{i+1}(\cdot)$, then $\pi_{i+1}(\cdot)$ is a stationary distribution for the iteration.

(See [36] for a proof). □

*Proposition 2*:

In the limit of $\Delta\vartheta \to 0$, inner iteration produces a chain that is reversible with respect to $\pi_{t_{i+1}}(\cdot)$.

*Proof*:

For $x = y$, this equality holds trivially. We consider when $x \neq y$. Specifically, keeping in mind the equation for the inner iteration as in Step 5 of the Algorithm, we interpret $X_{i+1,k} = x$ and $X_{i+1,k+1} = y$. Letting $\mathcal{I}$ to denote the indicator function, we may write

$$\pi_{i+1}(dx) = E_P[\mathcal{I}(x, x+dx) | \mathcal{F}_{i+1}^Y]$$

$$= E_Q[\mathcal{I}(x, x+dx)\exp(h(x)\Delta Y_i - \frac{1}{2}h^2(x)\Delta t)] \quad (A23)$$

$$= \exp(h(x)\Delta Y_i - \frac{1}{2}h^2(x)\Delta t)Q(dx)$$

Here $Q$ denotes a new measure, equivalent to $P$, such that $\Delta Y_i$ is rendered a zero-mean Brownian increment (i.e. the conditional expectation may be written as an unconditional one). Now define

$$\mathbf{P}_{i+1}(x, dy) = \frac{1}{\sqrt{2\pi c^2}}\exp((y - x - Gh(x)\Delta\vartheta)^2 / c^2)Q(dy) \quad (A24)$$

Here the transitional distribution $\mathbf{P}_{i+1}(x, \cdot)$ corresponds to the change of measure effected by the additive update via the inner iterate over $\vartheta$. Using A23 and A24, and letting $c^2 = \text{var}(\Delta Y_i)$, we get

$$\pi_{i+1}(dx)\mathbf{P}(x, dy) = \frac{1}{\sqrt{2\pi c^2}}\exp(h(x)\Delta Y_i - \frac{1}{2}h^2(x)\Delta t)Q(dx)\exp((y - x - Gh(x)\Delta\vartheta)^2 / c^2)Q(dy)$$

(A25)

$$\pi_{i+1}(dy)\mathbf{P}(y, dx) = \frac{1}{\sqrt{2\pi c^2}}\exp(h(y)\Delta Y_i - \frac{1}{2}h^2(y)\Delta t)Q(dy)\exp((y - x - Gh(y)\Delta\vartheta)^2 / c^2)Q(dx)$$

(A26)

where $h(y)$ may be Taylor-expanded as:

$$h(y) = h(x) + h'(x)(y - x) + (1/2)h''(x)(y - x)^2 + \ldots \quad (A27)$$

Note that $|y - x|$ is of order $\mathcal{O}(\Delta\vartheta)$, where $\Delta\vartheta_k = \beta_k > 0$. Replacing the expansion of $h(y)$ in A26 and ignoring all the terms that has order $\mathcal{O}(\Delta\vartheta)$ or higher, we get,

$$\pi_{i+1}(dx)\mathbf{P}_{i+1}(x, dy) = \pi_{i+1}(dy)\mathbf{P}_{i+1}(y, dx)$$

□

*Definition 2*:

For $\varepsilon > 0$, a subset $C \subseteq \Omega$ is small (or, $(n_0, \varepsilon, \nu)$-small) if there exists a positive integer $n_0$ and a probability measure $\nu(\cdot)$ on $\Omega$ such that the following minorization condition holds:

$$\mathbf{P}^{n_0}(x, \cdot) \geq \varepsilon \nu(\cdot)$$

i.e. $\mathbf{P}^{n_0}(x, A) \geq \varepsilon \nu(A)$ for all $x \in \Omega$ and all measurable $A \subseteq \Omega$.

*Proposition 3*:

Consider a $\vartheta$-chain with invariant probability distribution $\pi_{i+1}(\cdot)$. Suppose that, in the special case of $C = \Omega$ (i.e. the entire state space is small), the minorization condition holds for some $n_0 \in \mathbf{N}$, $\varepsilon > 0$ and probability measure $\nu(\cdot)$. Then the chain is uniformly ergodic and $\left\| \mathbf{P}^k(x, \cdot) - \pi(\cdot) \right\|_{tot} \leq (1 - \varepsilon)^{\lfloor k/n_0 \rfloor}$ for all $x \in \Omega$, where $\lfloor r \rfloor$ is the greatest integer not exceeding $r$. Here $\|\cdot\|_{tot}$ denotes the total variation distance.

*Proof:*

See [35; Theorem 2] for a proof. □

*Corollary 1*:

Suppose that the minorization condition holds for some $n_0 \in \mathbf{N}$, $\varepsilon > 0$ and probability measure $\nu(\cdot)$. Then,

$$\left\| \mathbf{P}^1(x, \cdot) - \mathbf{P}^\kappa(y, \cdot) \right\|_{tot} \leq (1 - \varepsilon)^{\lfloor 1/n_0 \rfloor} + (1 - \varepsilon)^{\lfloor \kappa/n_0 \rfloor}$$

Proof:

For some $n_0 \in \mathbf{N}$ and $\varepsilon > 0$

$$\left\| \mathbf{P}^1(x, \cdot) - \pi(\cdot) \right\|_{tot} \leq (1 - \varepsilon)^{\lfloor 1/n_0 \rfloor}$$

$$\left\| \mathbf{P}^\kappa(x, \cdot) - \pi(\cdot) \right\|_{tot} \leq (1 - \varepsilon)^{\lfloor k/n_0 \rfloor}$$

$$\left\|\mathbf{P}^1(x,\cdot)-\mathbf{P}^\kappa(x,\cdot)\right\|_{tot} \leq \left\|\mathbf{P}^1(x,\cdot)-\pi(\cdot)\right\|_{tot} + \left\|\mathbf{P}^\kappa(x,\cdot)-\pi(\cdot)\right\|_{tot}$$

$$\leq (1-\varepsilon)^{\lfloor 1/n_0 \rfloor} + (1-\varepsilon)^{\lfloor \kappa/n_0 \rfloor}$$

□

*Proposition 4*:

$$E_P\left[\left|\pi'^e_{i+1,1}(\phi)-\pi'^e_{i+1,\kappa}(\phi)\right|^2\right] \leq \left(\left\|\mathbf{P}^1(x,|\phi|^2)\right\| + \left\|\mathbf{P}^\kappa(x,|\phi|^2)\right\|\right) d_{Hell}(\mathbf{P}^1(x,\cdot),\mathbf{P}^\kappa(x,\cdot))$$

Proof:

We use the following property of the Hellinger metric:

$$d_{Hell}(\mathbf{P}^1(x,\cdot),\mathbf{P}^\kappa(x,\cdot)) \leq \sqrt{\left\|\mathbf{P}^1(x,\cdot)-\mathbf{P}^\kappa(x,\cdot)\right\|_{tot}}$$

Now we may write

$$\left(E_P\left[\left|\pi'^e_{i+1,1}(\phi)-\pi'^e_{i+1,\kappa}(\phi)\right|^2\right]\right)^{\frac{1}{2}} = \left(E_P\left[\left|\mathbf{P}^1(x,\phi)-\mathbf{P}^\kappa(x,\phi)\right|^2\right]\right)^{\frac{1}{2}}$$

$$\leq \left(\left\|\mathbf{P}^1(x,|\phi|^2)\right\| + \left\|\mathbf{P}^\kappa(x,|\phi|^2)\right\|\right)^{\frac{1}{2}} d_{Hell}(\mathbf{P}^1(x,\cdot),\mathbf{P}^\kappa(x,\cdot))$$

Using the bounds found above, we directly arrive at **Theorem 1**. □

### References

1. Oksendal, B.K. (2003), *Stochastic Differential Equations-An Introduction with Applications*, 6th ed., Springer, New York.

2. Kalman, R. E. and Bucy, R. S. (1961), New results in linear filtering and prediction theory, *Trans. ASME, Ser. D, J. Basic Eng.* 83: 95–107.

3. Jazwinski, A. H. (1970), *Stochastic Processes and Filtering Theory*, Academic Press, New York.

4. Julier, S. J. and Uhlmann, J. K. (1997), A New Extension of the Kalman Filter to Nonlinear Systems, *In Proc. of AeroSense: The 11th Int. Symp. On Aerospace/Defence Sensing, Simulation and Controls*.

5. Doucet, A., Godsill, S. and Andrieu, C. (2000), On sequential Monte Carlo sampling methods for Bayesian filtering, *Statistics and Computing* 10:197–208.

6. Arulampalam, S., Maskell, N., Gordon, N. and Clapp, T. (2002), A tutorial on particle filters for online nonlinear/non-Gaussian Bayesian tracking, *IEEE Transactions on Signal Processing* 50:174-188.

7. Gordon, N. J., Salmond, D. J. and Smith, A. F. M. (1993), Novel approach to nonlinear/non-Gaussian Bayesian state estimation, *Radar and Signal Processing, IEE Proceedings F* 140**:**107-113.

8. Budhiraja, A., Chen, L. and Lee, C. (2007), A survey of numerical methods for nonlinear filtering problems, *Physica D* 230:27-36.

9. Snyder, C., Bengtsson, T., Bickel, P. and Anderson, J. (2008), Obstacles to High-Dimensional Particle Filtering, *Mon. Wea. Rev.* 136**:**4629–4640.

10. Chorin, A. J. and Tu, X. (2009), Implicit sampling for particle filters, *Proc. Natl. Acad. Sci. USA* 106: 17249-17254.

11. Jeroen, D. H., Thomas, B.S., Gustafsson, F. (2006), On Resampling Algorithms For Particle Filters, *IEEE Nonlinear Statistical Signal Processing Workshop* 79-82, Cambridge, UK.

12. Geweke, J. and Tanizaki, H. (1999), On Markov Chain Monte Carlo methods for nonlinear and non-gaussian state-space models, *Commun. Stat. Simul. C* 28:867–894.

13. Budhiraja, A. and Kallianpur, G. (1996), Approximations to the solutions of the Zakai equation using multiple Wiener and Stratonovich integral expansions, *Stochastics and Stochastic Reports* 56:271-315.

14. Di Masi, G .B., Pratelli, M. and Runggaldier, W. J. (1985), An approximation for the nonlinear filtering problem with error bound, *Stochastics* 14(4):247-271.

15. Elliott, R. J. and Glowinski, R. (1989), Approximations to solutions of the Zakai filtering equation, *Stochastic Analysis and Applications* 7(2):145-168.

16. Florschinger, P. and Le Gland, F. (1991), Time discretization of the Zakai equation for diffusion processes observed in correlated noise, *Stochastics and Stochastics Reports* 35:233-256.

17. Budhiraja, A. and Kallianpur, G. (1997) The Feynman-Stratonovich Semigroup and Stratonovich Integral Expansions in Nonlinear Filtering, *Appl. Math. Optim.* 35:91-116.

18. Gobet, E., Pagès,G., Pham, H. and Printems, J. (2006), Discretization and simulation of the Zakai equation, *SIAM J. Numer. Anal.* 44(6):2505-2538.

19. Ahmed, N. U. and Radaideh, S. M. (1997), A powerful numerical technique solving Zakai equation for nonlinear filtering, *Dynamics and Control* 7:293-308.

20. Ito, K. and Rozovskii, B. (2000), Approximation of the Kushner equation for nonlinear filtering, *SIAM J. Control Optim.* 38(3):893-915.

21. Milstein, G. V. and Tretyakov, M.V. (2009), Solving parabolic stochastic partial differential equations via averaging over characteristics, *Math. Comp.* 78:2075–2106.

22. Crisan, D. and Lyons, T. (1999) A particle approximation of the solution of the Kushner-Stratonovich equation, *Probab. Theory Related Fields* 115:549-578.

23. Ionides, E. L., Bhadra,A., Atchadé, Y. and King, A. (2011), Iterated Filtering, *The Annals of Statistics* 39(3):1776-1802.

24. Roy, D., Saha, N. and Dash, M. K. (2008), Weak Forms of the Locally Transversal Linearization (LTL) Technique for Stochastically Driven Nonlinear Oscillators, *Appl. Math. Model.* 32:1657-1681.

25. Saha, N. and Roy, D. (2007) Higher order weak linearizations of stochastically driven nonlinear oscillators, *Proc. R. Soc. A* 463:1827–1856.

26. Kirkpatrick, S., Gelatt, C. D. and Vecchi, M. P. (1983) Optimization by simulated annealing, *Sci.* 220:671-680.

27. Lam, J. and Delosme, J.-M. (1988), An efficient simulated annealing schedule: derivation, *Technical Report 8816 Yale Electrical Engineering Department, New Haven, Connecticut.*

28. Justel, A., Peña, D. and Zamar, R. (1997) A multivariate Kolmogorov-Smirnov test of goodness of fit, *Statistics & Probability Letters* 35:251-259.

29. Crisan, D. and Doucet, A. (2002), A survey of convergence results on particle filtering methods for practitioners, *IEEE Transactions on Signal Processing* 50(3).

30. Pitt, M. and Shephard, N. (1999), Filtering via simulation: Auxiliary particle filters, *J. Amer. Statist. Assoc.* 94:590–599.

31. Livings, D. M., Dance, S. L., and Nichols, N. K. (2008), Unbiased ensemble square root filters, *Physica D*, 237:1021-1028.

32. Roy, D. (2001), A new numeric-analytical principle for nonlinear deterministic and stochastic dynamical systems, *Proc. Roy. Soc. Lond. A* 457:539–566.

33. Roy, D. (2004), A family of lower- and higher-order transversal linearization techniques in non-linear stochastic engineering dynamics, *Int. J. Numer. Meth. Eng.* 61:764–790.

34. Kurtz, T. G. and Ocone, D. L., 1998, *Unique Characterization of Conditional Distributions in Nonlinear Filtering*, Ann. Prob., Vol. 16(1), pp. 80-107.

35. Kloeden, P. and Platen, E. (1992) *Numerical Solutions of Stochastic Differential Equations,* Springer, Berlin

36. Roberts, G. O. and Rosenthal J. S. (2004), General state space Markov chains and MCMC algorithms, Prob. Surveys 1:20-71.

Figures

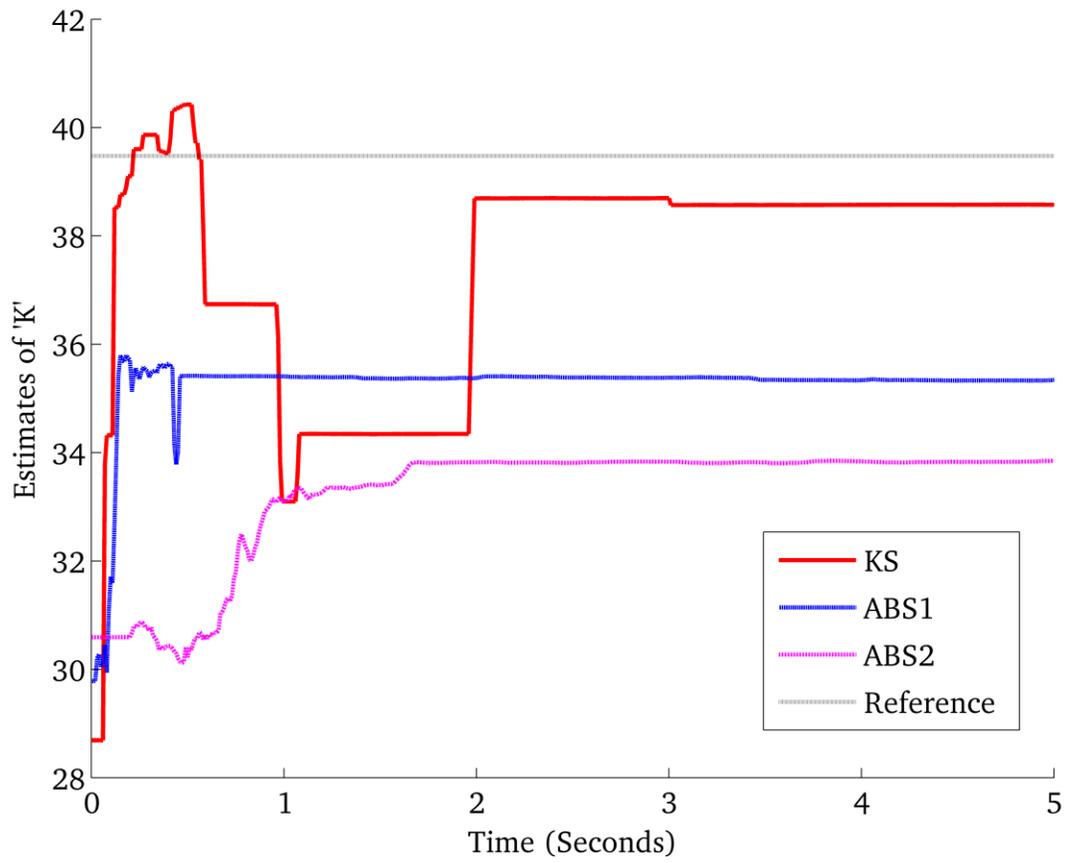

Figure 3.1 (a): Estimates of *k*

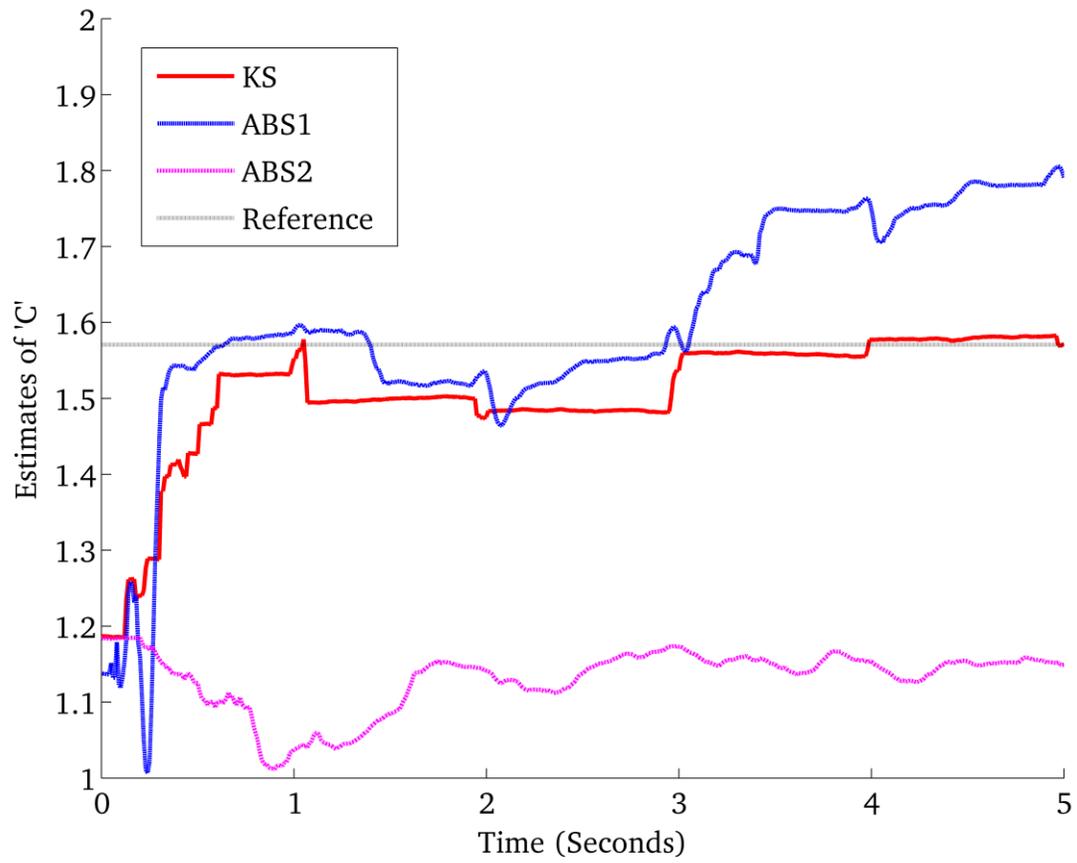

Figure 3.1 (b): Estimates of $c$

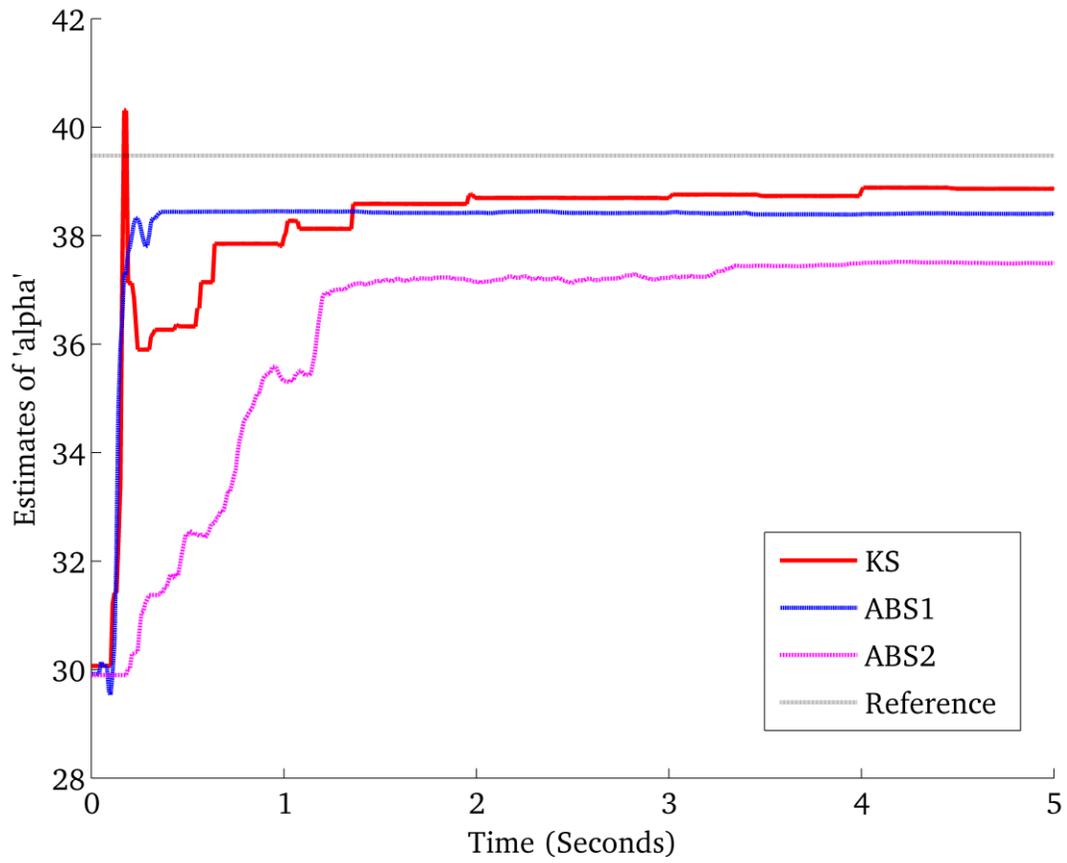

Figure 3.1 (c): Estimates of $\alpha$

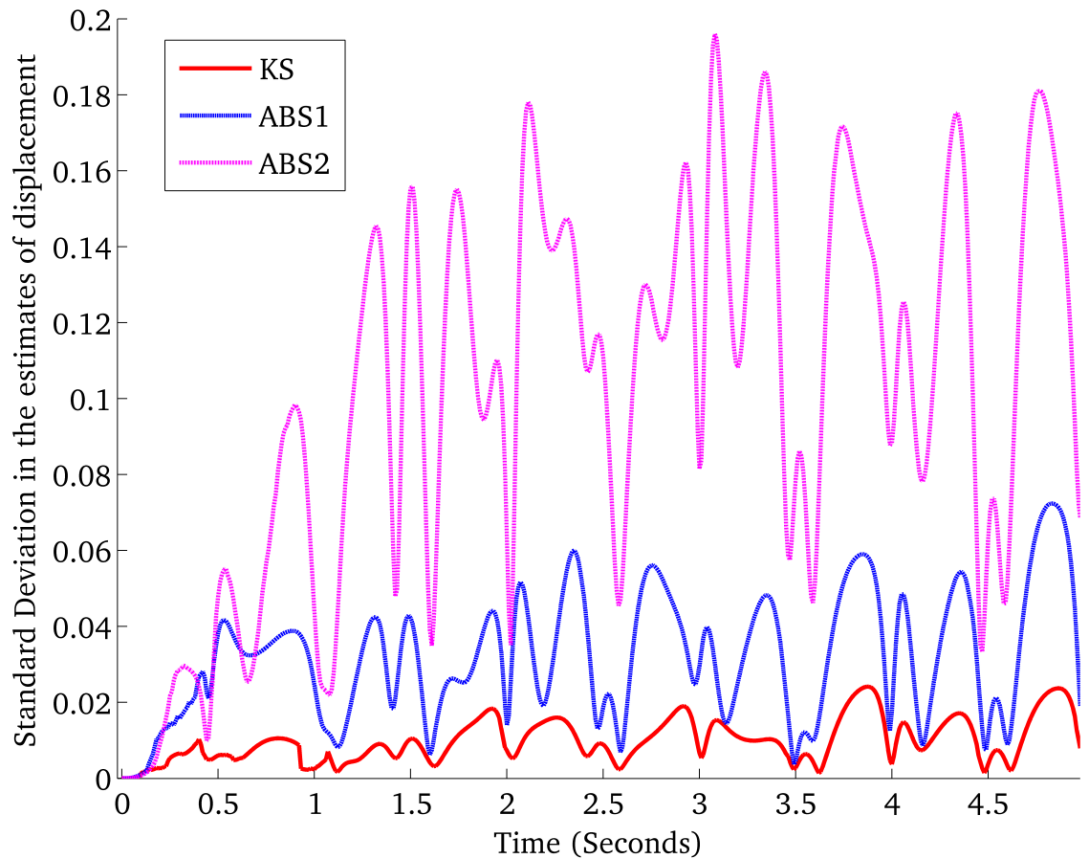

Figure 3.2(a): Standard deviation in the estimates of displacement

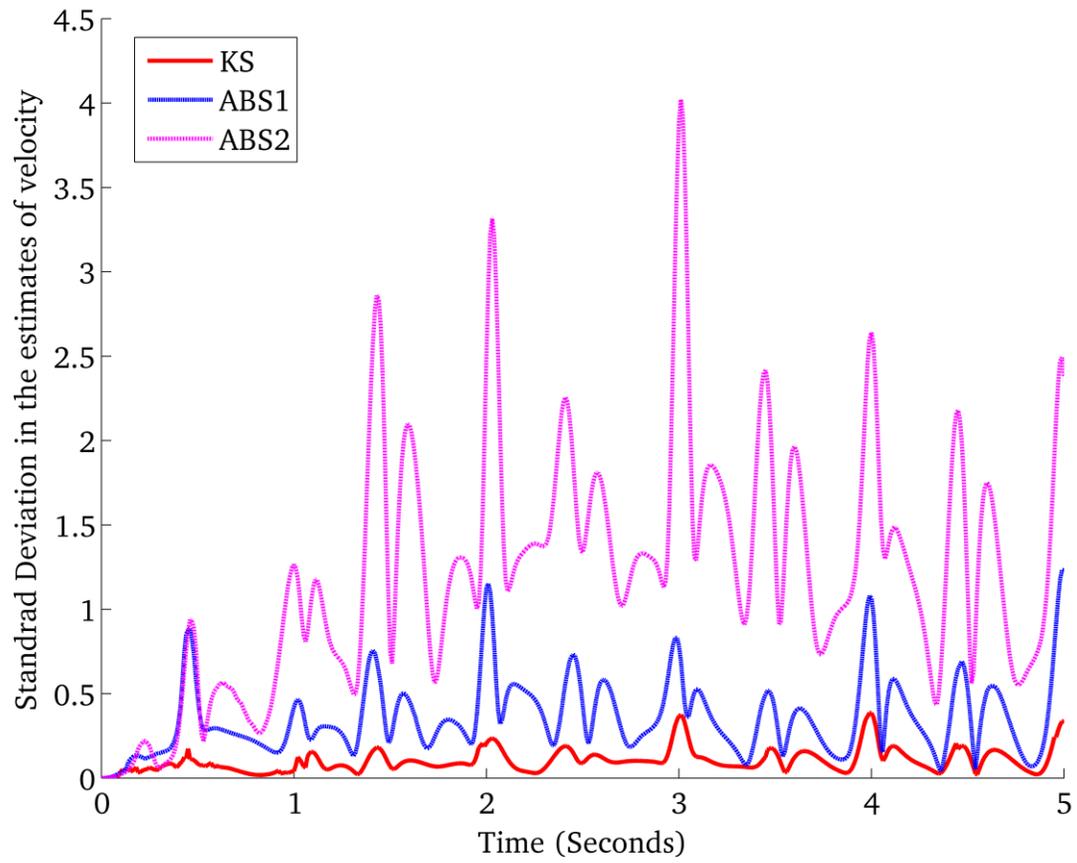

Figure 3.2(b): Standard deviation in the estimates of velocity

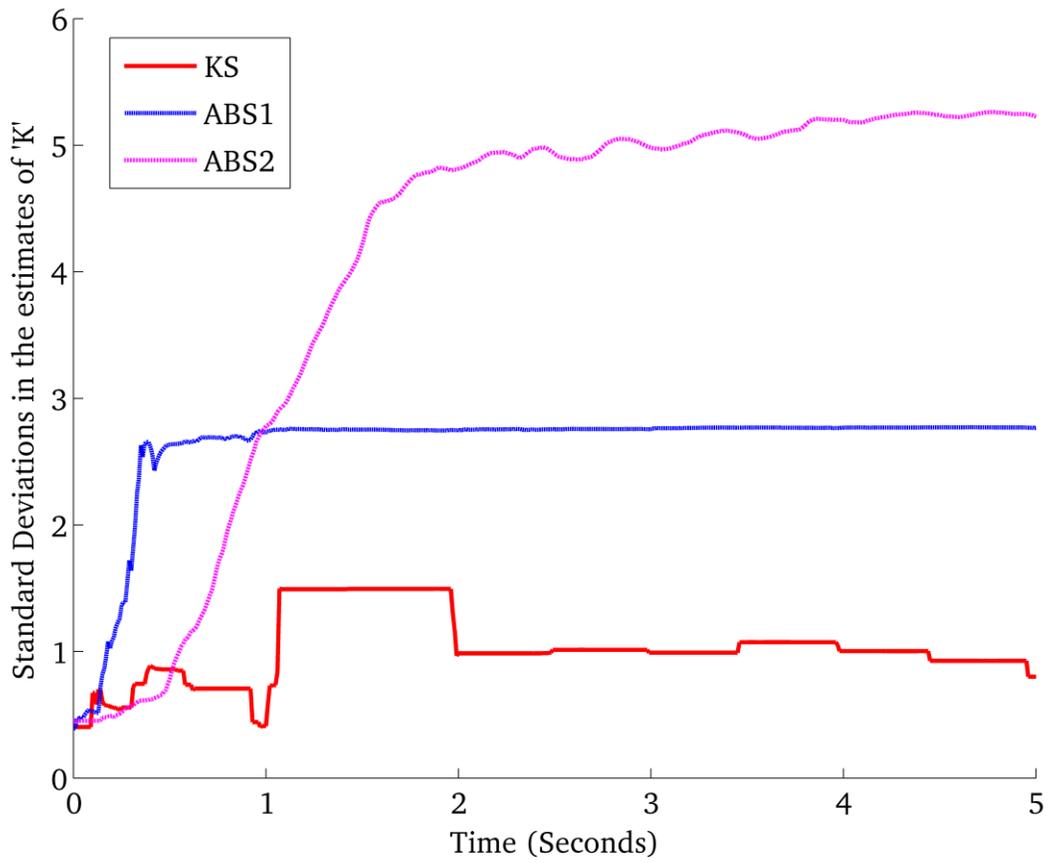

Figure 3.2(c): Standard deviation in the estimates of *k*

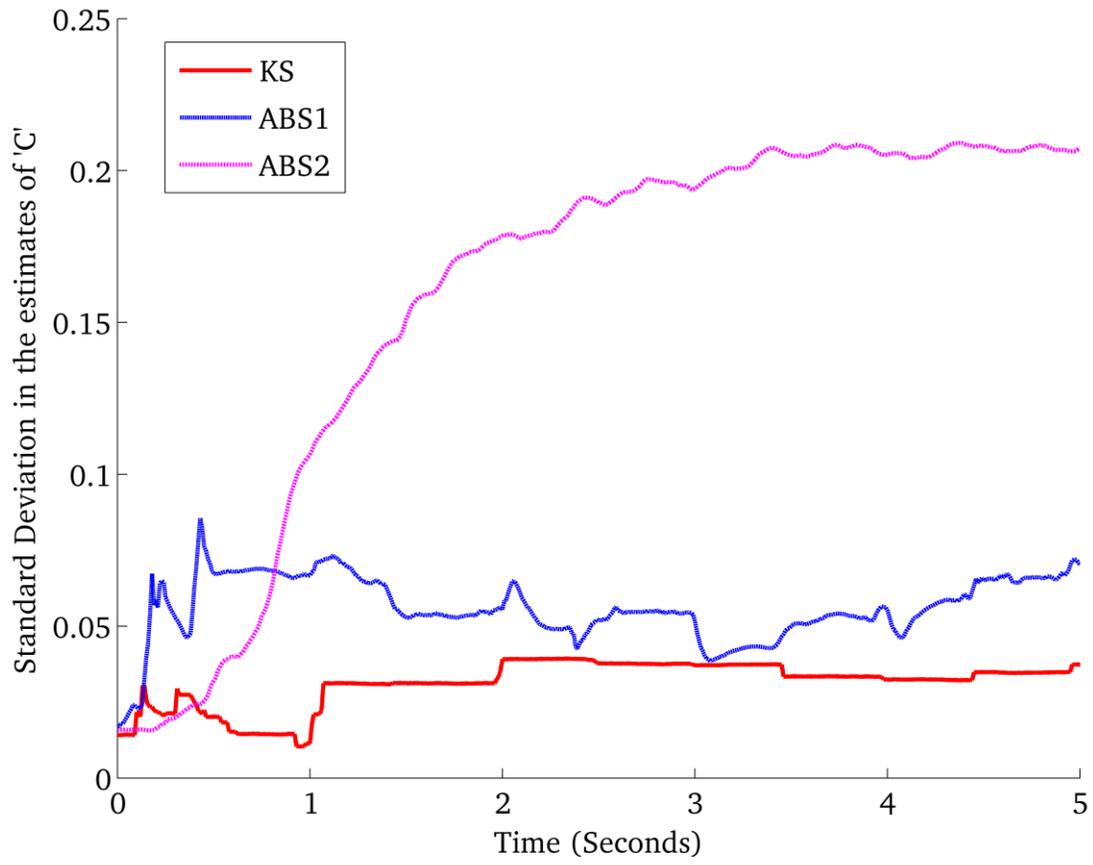

Figure 3.2 (d): Standard deviation in the estimates of $c$

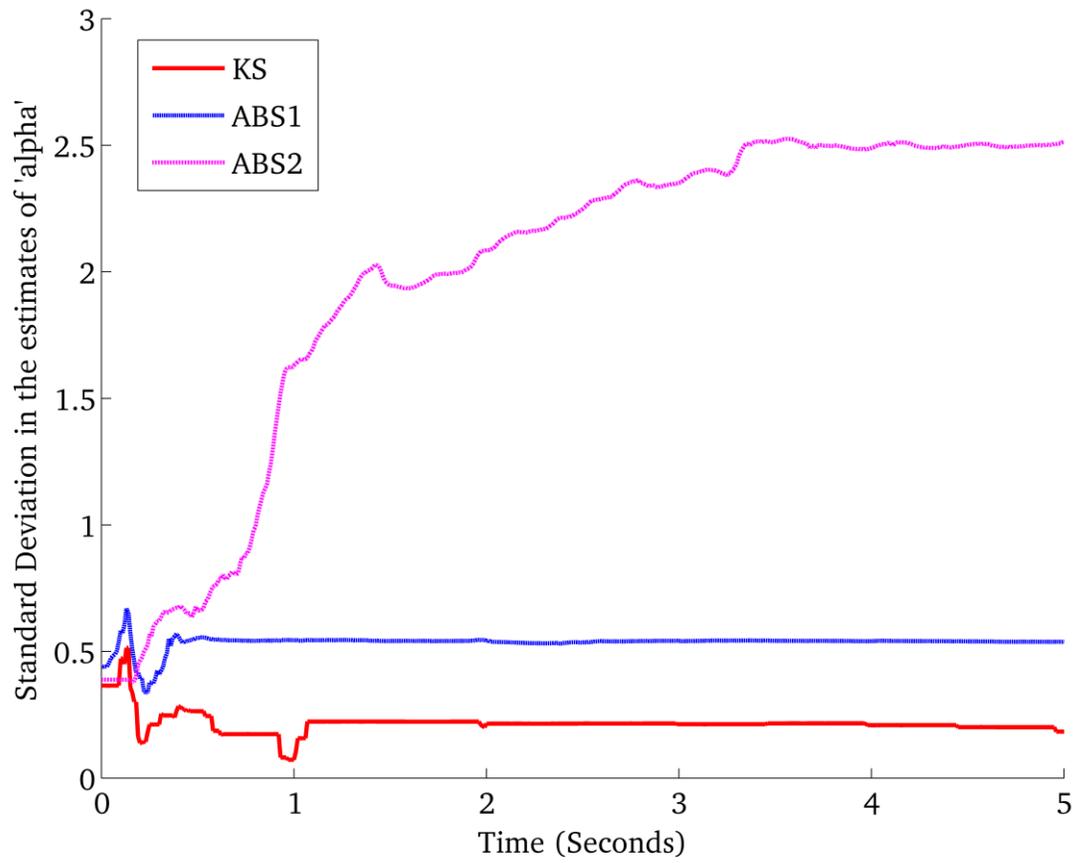

Figure 3.2 (e): Standard deviation in the estimates of $\alpha$

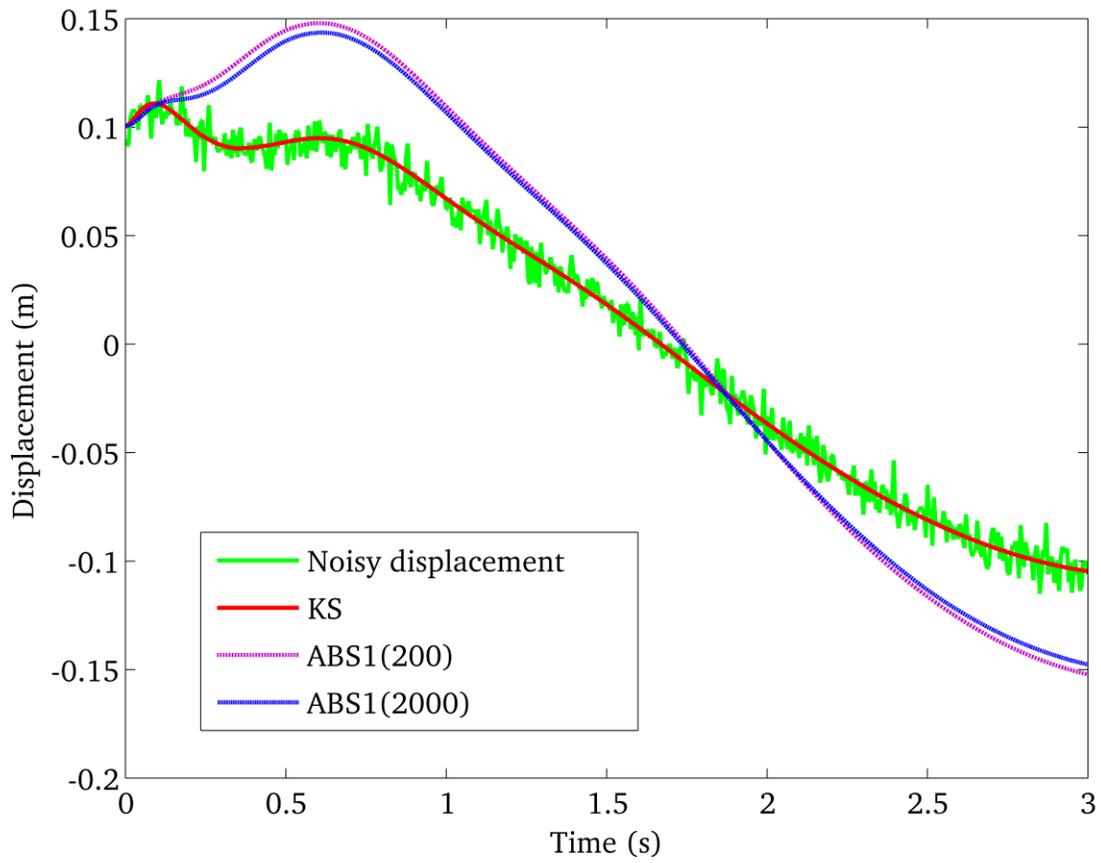

Figure 3.3 (a): Estimates of displacement of 4th floor

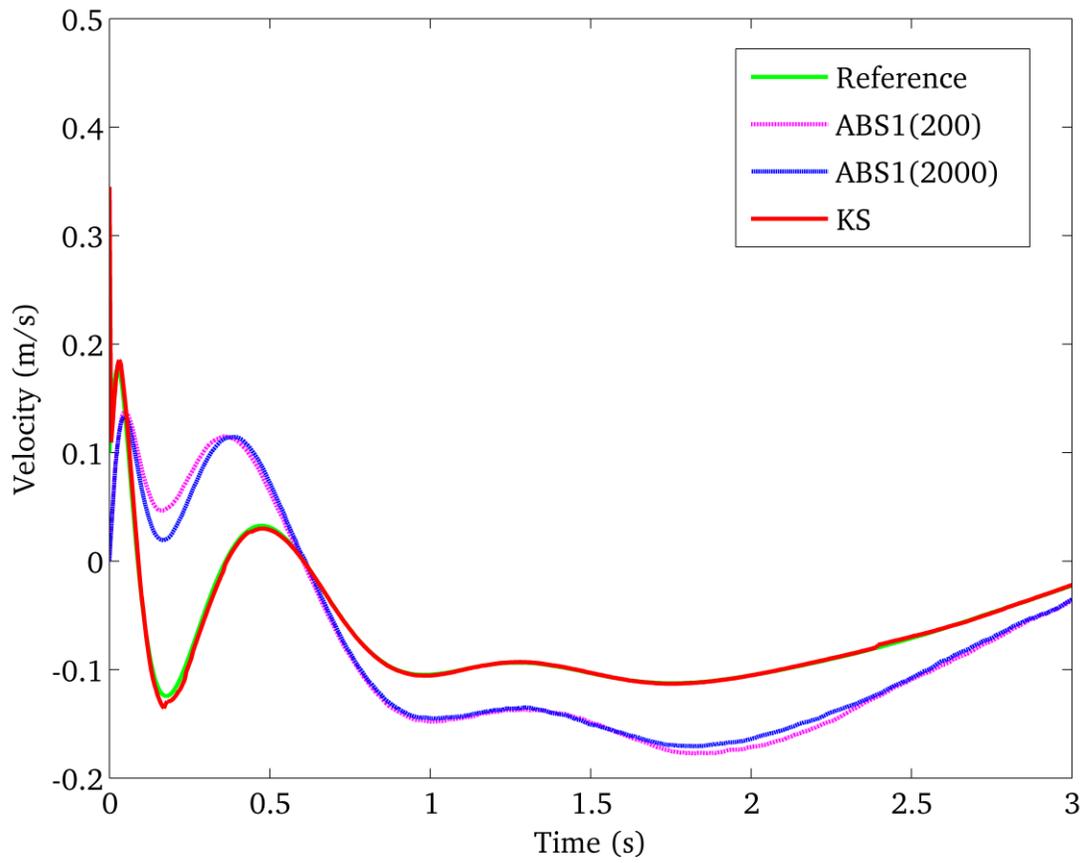

Figure 3.3 (b): Estimates of velocity of 4[th] floor

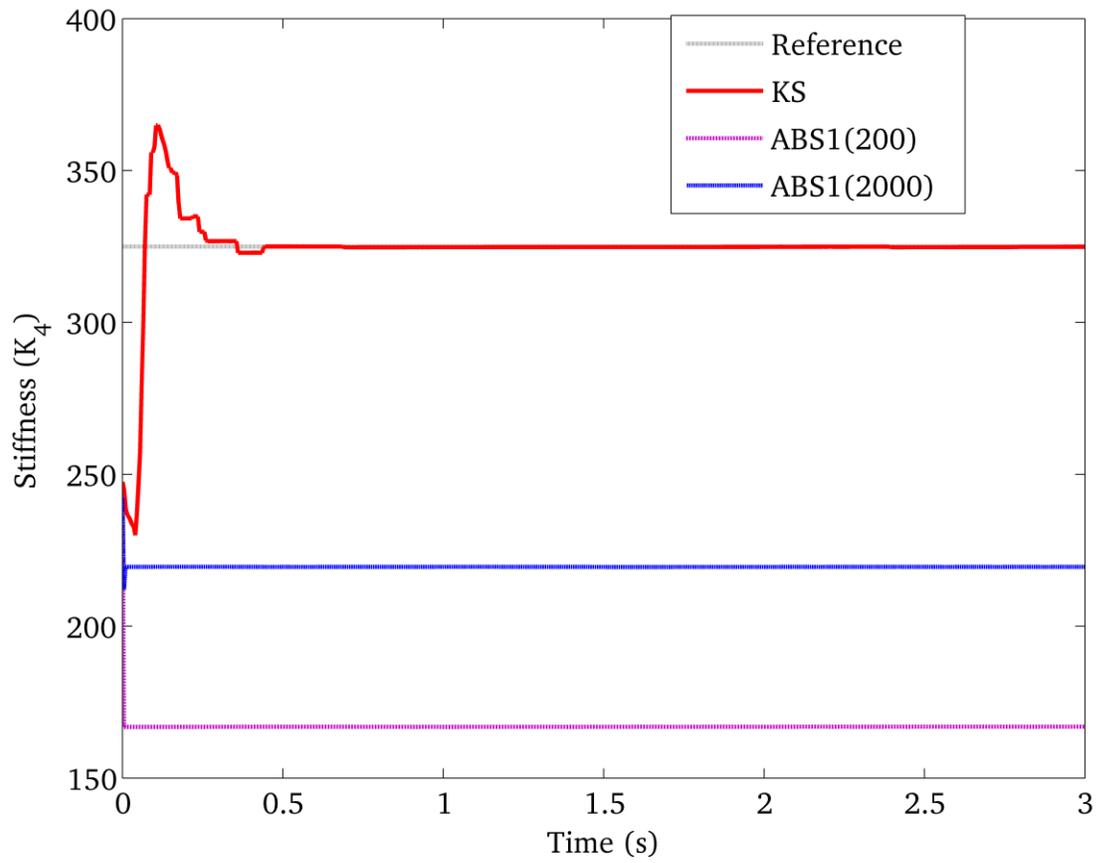

Figure 3.3 (c): Estimates of stiffness of 4$^{th}$ floor

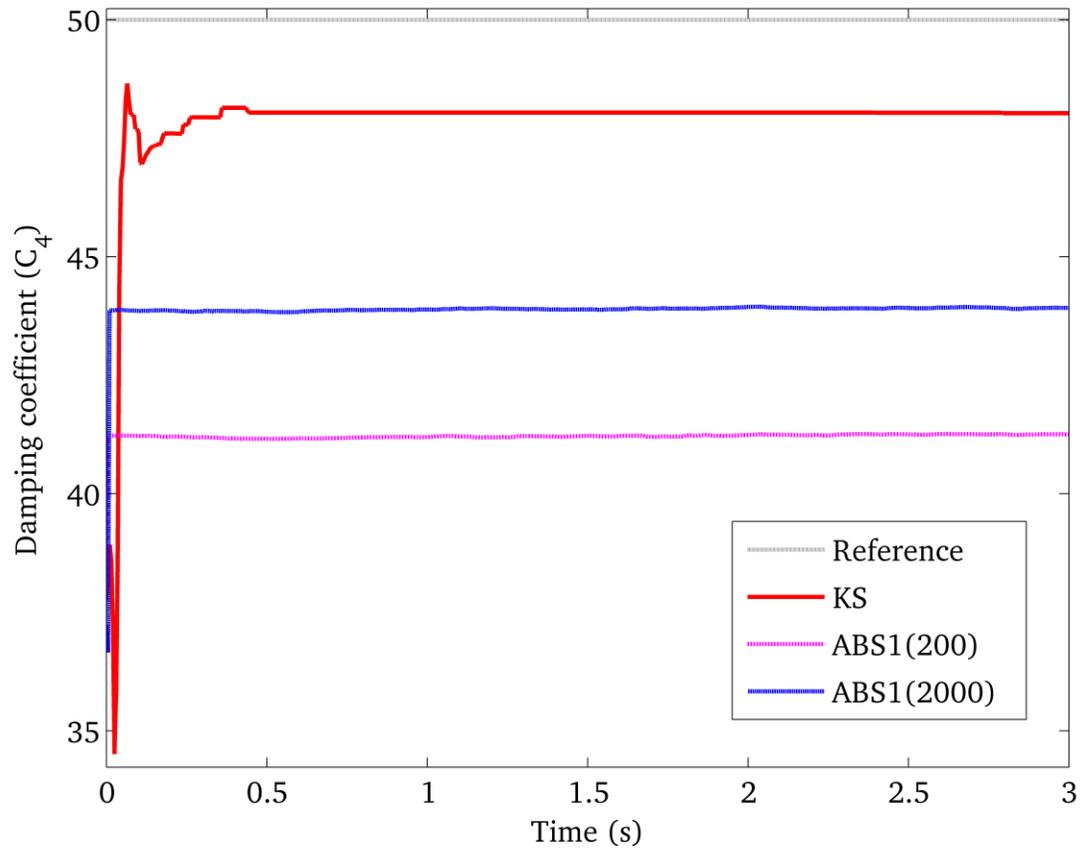

Figure 3.3 (d): Estimates of damping coefficient of $4^{th}$ floor

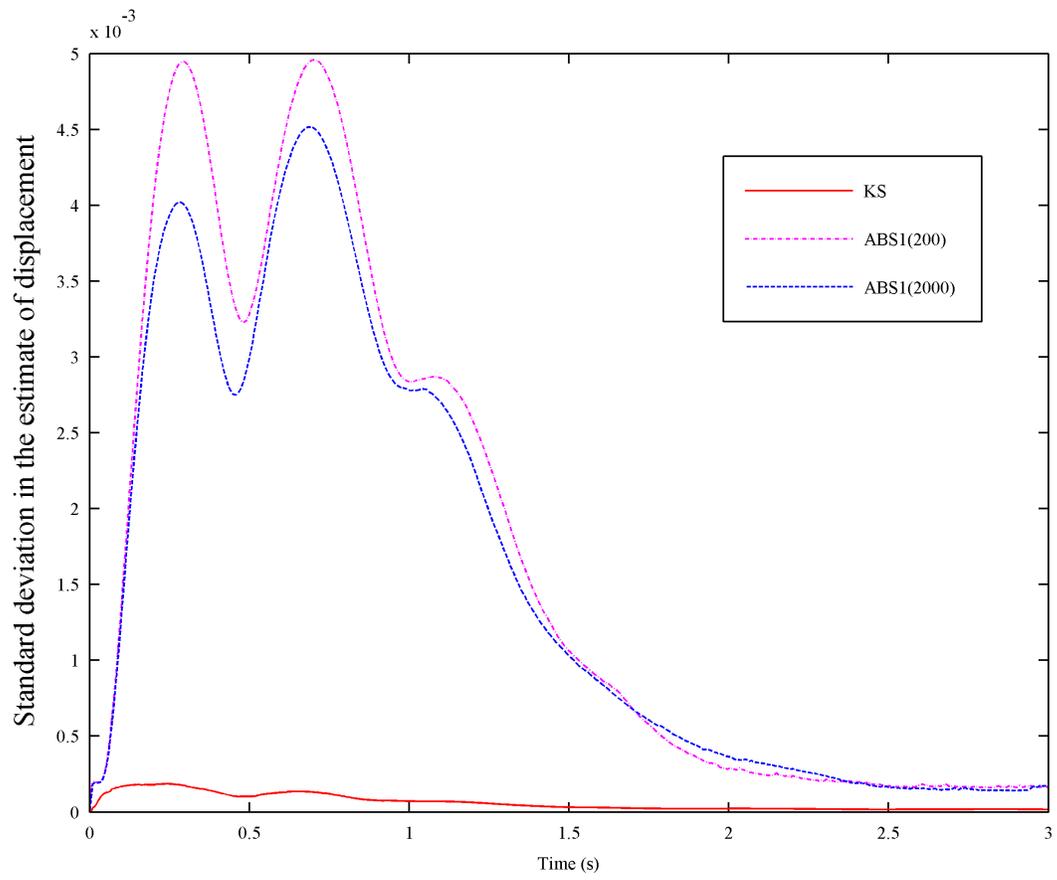

Figure 3.4 (a): Standard deviation in the estimates of 4$^{th}$ floor displacement

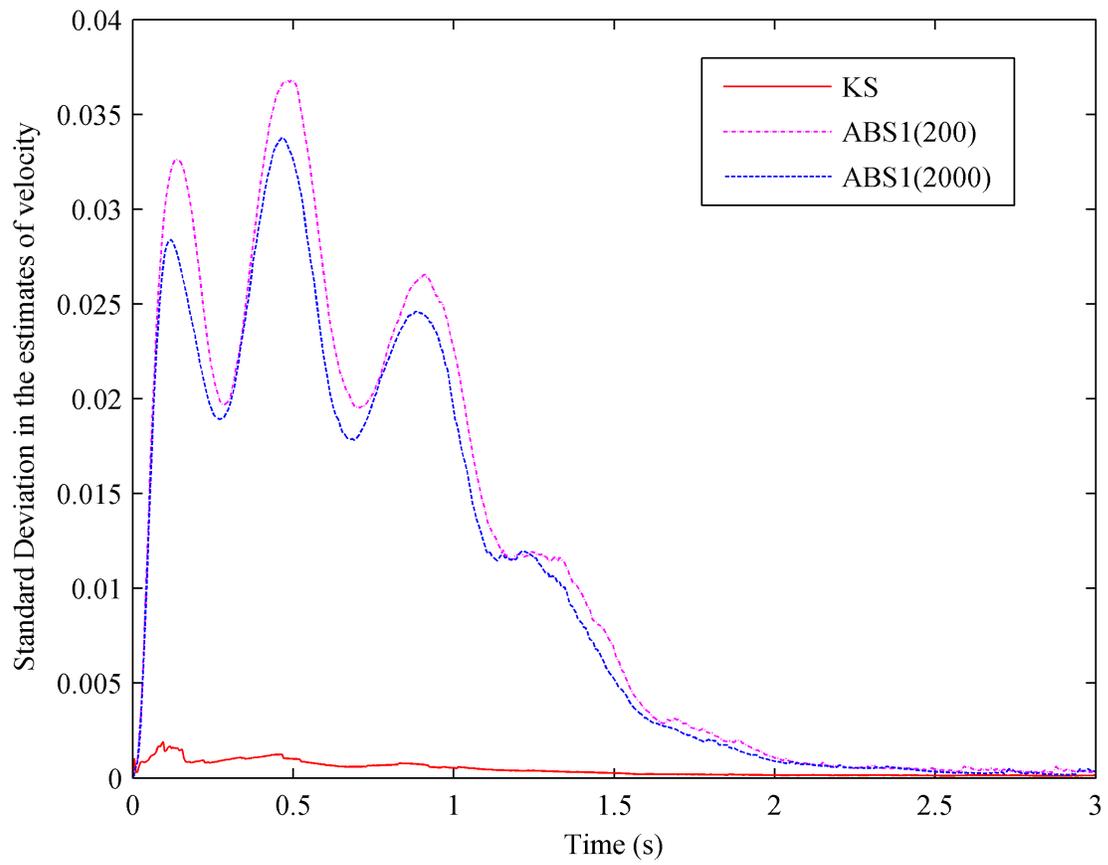

Figure 3.4 (b): Standard deviation in the estimates of 4$^{th}$ floor velocity

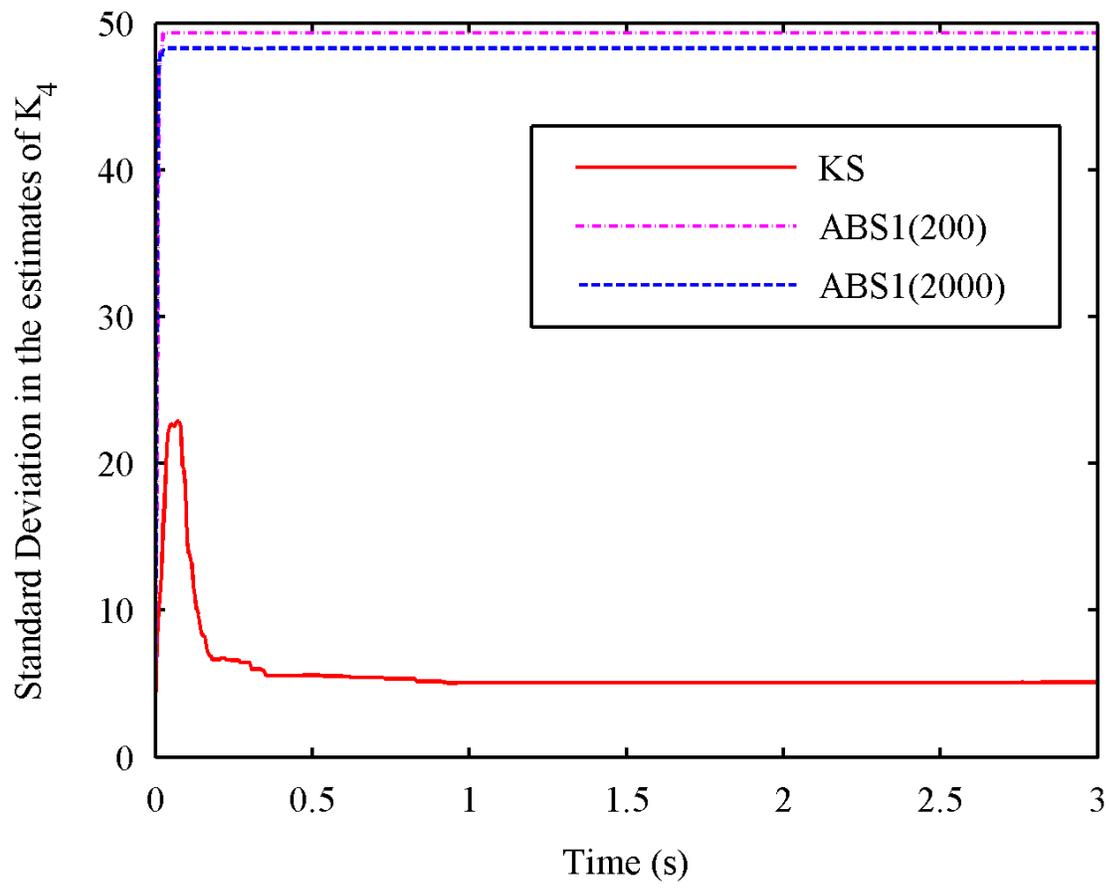

Figure 3.4 (c): Standard deviation in the estimates of stiffness of $4^{th}$ floor

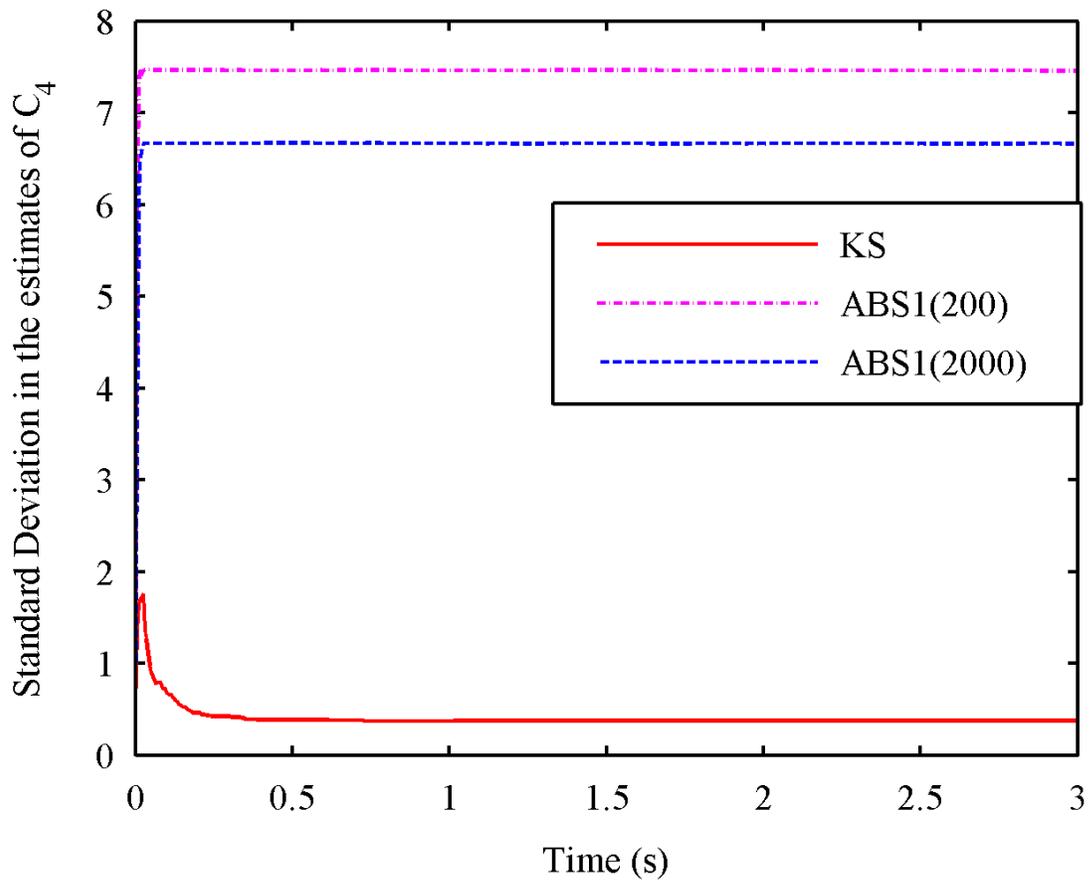

Figure 3.4 (d): Standard deviation in the estimates of damping coefficient of $4^{th}$ floor

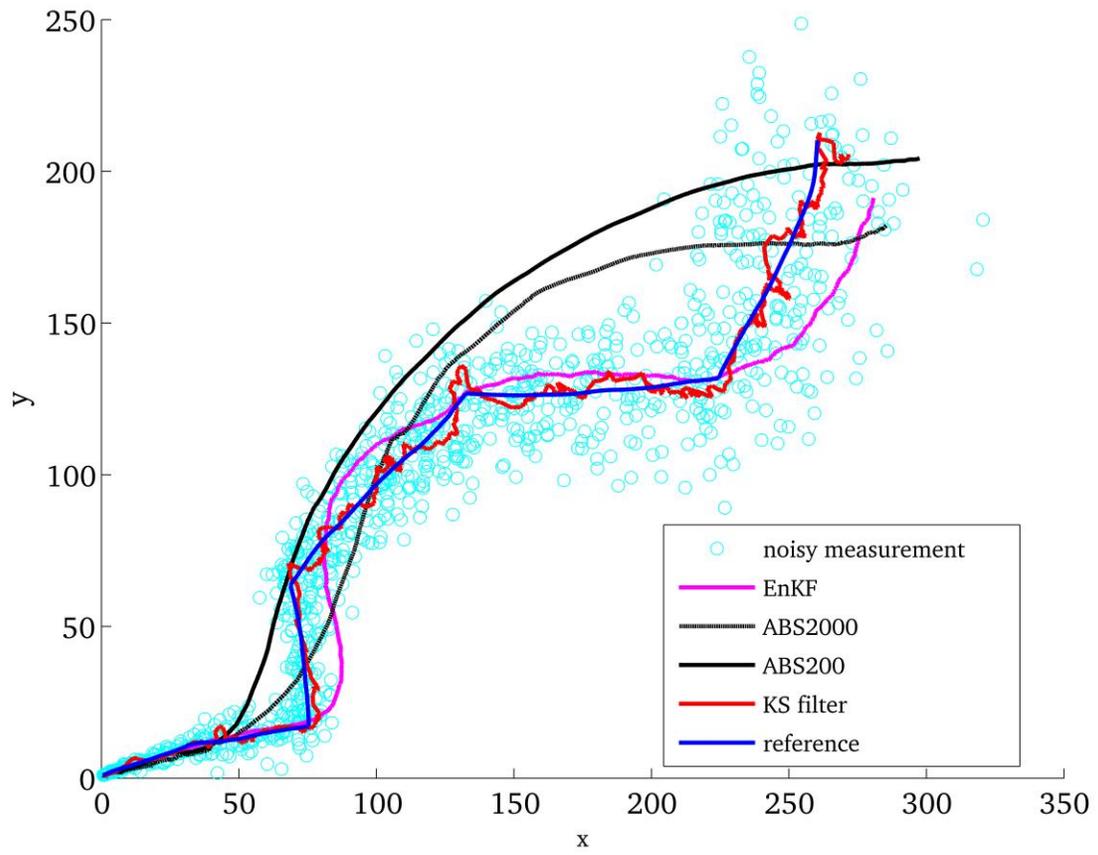

Figure 3.5: Ship trajectories estimated by various filters

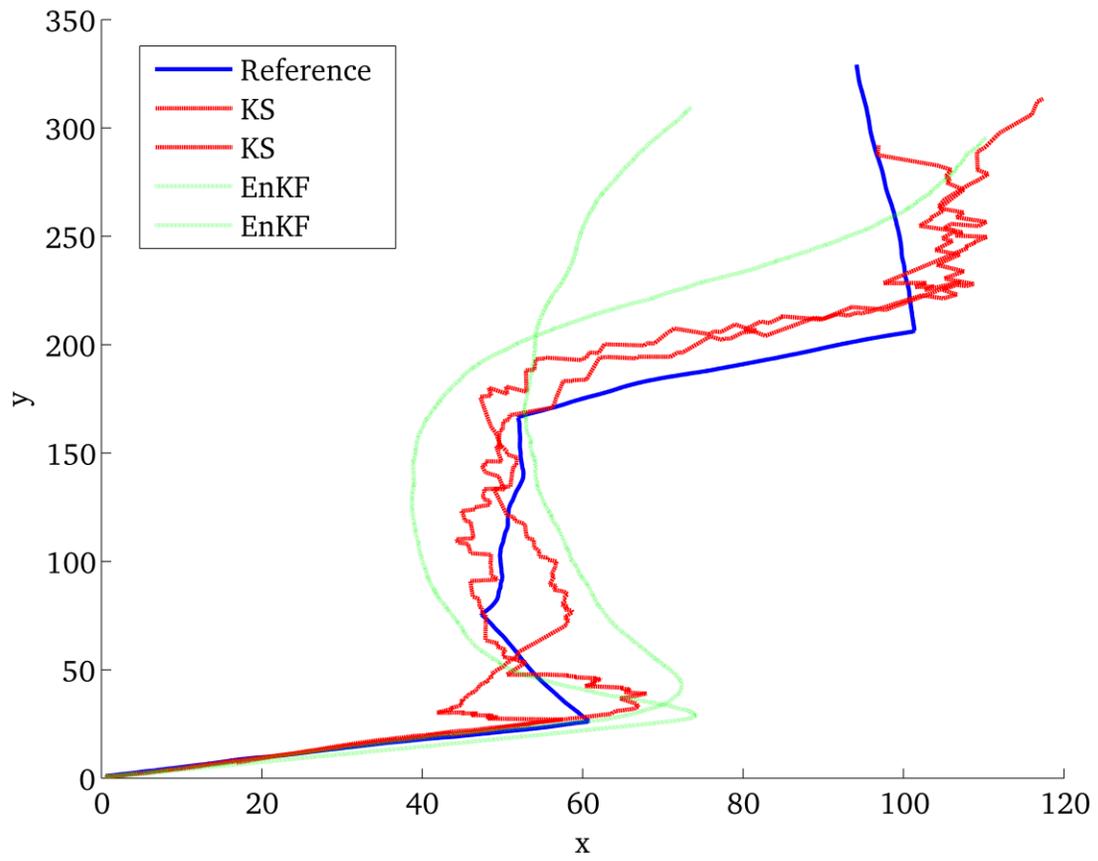

Figure 3.6: Ship trajectories estimated by KS and EnKF with $N = 5$ in two filter runs